\documentclass[conference]{IEEEtran}
\IEEEoverridecommandlockouts
\usepackage{cite}
\usepackage{amsmath,amssymb,amsfonts}
\usepackage{algorithmic}
\usepackage{graphicx}
\usepackage{textcomp}
\usepackage{xcolor}
\def\BibTeX{{\rm B\kern-.05em{\sc i\kern-.025em b}\kern-.08em
    T\kern-.1667em\lower.7ex\hbox{E}\kern-.125emX}}

\usepackage{tabulary}
\usepackage{booktabs}
\usepackage{hyperref}
\usepackage{lipsum}
\usepackage{pifont}
\usepackage{stmaryrd}
\usepackage{paralist}
\usepackage{enumitem}
\usepackage{xspace}
\usepackage{dsfont}
\usepackage{subcaption}
\newcommand{\filler}[1]{\relax}
\usepackage{comment}

\includecomment{techrep}

\usepackage{cleveref}
\definecolor{OwnAzure}{HTML}{336699}
\definecolor{OwnCerulean}{HTML}{CAE2FE}
\definecolor{OwnOliveGreen}{HTML}{556B2F}
\definecolor{KamPurple}{HTML}{907C97}
\usepackage[colorinlistoftodos,prependcaption,textsize=tiny]{todonotes}%
\usepackage{xargs}                      %

\newcommandx{\todolarge}[2][1=]{\todo[inline,size=\large,linecolor=OwnAzure,backgroundcolor=OwnCerulean,bordercolor=OwnAzure,#1]{#2}}

\newcommandx{\todoai}[2][1=]{\todo[inline,linecolor=OwnAzure,backgroundcolor=OwnCerulean,bordercolor=OwnAzure,#1]{#2}}

\newcommand{\todojd}[1]{}
\newcommand{\jesseinline}[1]{}

\newcommandx{\addref}[2][1=]
{\todo[inline,linecolor=blue,backgroundcolor=blue!50,bordercolor=blue,#1]{Add reference. #2}}
\newcommandx{\unsure}[2][1=]{\todo[inline, linecolor=red,backgroundcolor=red!25,bordercolor=red,#1]{#2}}
\newcommandx{\change}[2][1=]{\todo[inline, linecolor=blue,backgroundcolor=blue!25,bordercolor=blue,#1]{#2}}
\newcommandx{\info}[2][1=]{\todo[linecolor=OwnOliveGreen,backgroundcolor=OwnOliveGreen!25,bordercolor=OwnOliveGreen,#1]{#2}}
\newcommandx{\improvement}[2][1=]{\todo[linecolor=Plum,backgroundcolor=Plum!25,bordercolor=Plum,#1]{#2}}
\newcommandx{\thiswillnotshow}[2][1=]{\todo[disable,#1]{#2}}

\definecolor{darkred}{rgb}{0.5,0,0}
\definecolor{darkgreen}{rgb}{0,0.5,0}
\definecolor{darkblue}{rgb}{0,0,0.5}

\graphicspath{{images/}}

\newcommand{\sys}{Servo\xspace}

\usepackage{glossaries}
\newacronym{mmog}{MMOG}{massively multiplayer online game}

\newacronym{minecraftlike}{MVE}{modifiable virtual environment}
\newacronym{mcgame}{MVE}{modifiable virtual environment}
\newcommand{\mve}{\gls*{mcgame}\xspace}
\newcommand{\mvepl}{\glspl*{mcgame}\xspace}
\newcommand{\mves}{\mvepl}

\newacronym{faas}{FaaS}{Function-as-a-Service}
\newcommand{\faas}{\gls*{faas}\xspace}
\newacronym{baas}{FaaS}{Backend-as-a-Service}

\newacronym{mmo}{MMO}{Massively Multiplayer Online Game}
\newcommand{\mmo}{\gls*{mmo}\xspace}

\newcommand{\mmos}{\glspl*{mmo}\xspace}

\newacronym{qos}{QoS}{quality of service}
\newcommand{\qos}{\gls*{qos}\xspace}

\newacronym{qoe}{QoE}{quality of experience}

\newacronym{rms}{RMS}{resource management and scheduling}
\newcommand{\rms}{\gls*{rms}\xspace}

\newcommand{\das}{DAS-5\xspace}

\newacronym{aws}{AWS}{Amazon Web Services}
\newcommand{\aws}{\gls*{aws}\xspace}

\newacronym{dc}{dyconit}{dynamic consistency unit}
\newacronym{aoi}{AoI}{area of interest}
\newacronym{is}{IS}{interest set}
\newglossaryentry{solution}{
	name={Dyconit},
	description={}
}
\newacronym{npc}{NPC}{Non-Playable Character}

\newcommand{\designref}[1]{%
	\begin{tikzpicture}[baseline=(char.base)]
		\node[draw=white,circle,inner sep=0.5pt, fill=gray, text=white] (char){\small #1};
	\end{tikzpicture}%
}

\newcommand{\modelref}[1]{%
	\protect\begin{tikzpicture}[baseline=(char.base)]
		\protect\node[draw=gray,circle,inner sep=0.5pt, fill=white, text=gray] (char){\small #1};
	\end{tikzpicture}%
}

\newcommand{\ocgithub}{\url{https://github.com/atlarge-research/opencraft}}

\newcommand{\vcutS}{\vspace*{-0.15cm}}
\newcommand{\vcutM}{\vspace*{-0.25cm}}

\makeatletter
\newcommand{\linebreakand}{%
  \end{@IEEEauthorhalign}
  \hfill\mbox{}\par
  \mbox{}\hfill\begin{@IEEEauthorhalign}
}
\makeatother

\begin{document}

\title{\sys: Increasing the Scalability of Modifiable Virtual Environments Using Serverless Computing
}

\author{
	[Technical Report on the ICDCS homonym article]
	\linebreakand
	\IEEEauthorblockN{Jesse Donkervliet}
	\IEEEauthorblockA{\textit{Vrije Universiteit Amsterdam} \\
		Amsterdam, Netherlands \\
		J.J.R.Donkervliet@vu.nl}
	\and
	\IEEEauthorblockN{Javier Ron}
	\IEEEauthorblockA{\textit{KTH Royal Institute of Technology} \\
		Stockholm, Sweden \\
		JavierRo@kth.se}
	\and
	\IEEEauthorblockN{Junyan Li}
	\IEEEauthorblockA{\textit{Vrije Universiteit Amsterdam} \\
		Amsterdam, Netherlands \\
		Junyan.Li@student.vu.nl}
	\linebreakand
	\IEEEauthorblockN{Tiberiu Iancu}
	\IEEEauthorblockA{\textit{Vrije Universiteit Amsterdam} \\
		Amsterdam, Netherlands \\
		T.Iancu@student.vu.nl}
	\and
	\IEEEauthorblockN{Cristina L. Abad}
	\IEEEauthorblockA{\textit{ESPOL} \\
		Guayaquil, Ecuador \\
		CAbad@fiec.espol.edu.ec}
	\and
	\IEEEauthorblockN{Alexandru Iosup}
	\IEEEauthorblockA{\textit{Vrije Universiteit Amsterdam} \\
		Amsterdam, Netherlands \\
		A.Iosup@vu.nl}
}

\maketitle

\begin{abstract}
	Online games with \textit{\mves} have become highly popular over the past decade. Among them, Minecraft---supporting hundreds of millions of users---is the best-selling game of all time, and is increasingly offered as a service.
Although Minecraft is architected as a distributed system, in production it achieves this scale by partitioning small groups of players over isolated game instances.
From the approaches that can help other kinds of virtual worlds scale, none is designed to scale \mves{}, which pose a unique challenge---a mix between the count and complexity of active in-game constructs, player-created in-game programs, and strict quality of service.
Serverless computing emerged recently and focuses, among others, on service scalability. Thus,
addressing this challenge, in this work we explore using serverless computing to improve \mve scalability.
To this end, we design, prototype, and evaluate experimentally \sys, a serverless backend architecture for \mves.
We implement \sys as a prototype and evaluate it using real-world experiments on two commercial serverless platforms, of Amazon Web Services~(AWS) and Microsoft Azure.
Results offer strong support that our serverless \mve can significantly increase the number of supported players per instance without performance degradation, in our key experiment
by 40 to 140 players per instance,
which is a significant improvement over state-of-the-art commercial and open-source alternatives.
We release \sys as open-source, on Github: \ocgithub. %

\end{abstract}

\begin{IEEEkeywords}
	MVE, serverless, scalability, virtual environments, online games,
	backend, design, experimentation %
\end{IEEEkeywords}

\glsresetall

\section{Introduction}
\label{sec:introduction}

Today's digital societies increasingly leverage online games, which are cloud-deployed services architected as distributed systems,
for other tasks with societal impact such as education~\cite{DBLP:conf/fdg/Bar-ElR20}, professional training~\cite{Melchiorre2022}, and social activism. %
Facilitating these use-cases requires support from a scalable \emph{virtual world}.
\emph{\Glspl*{mcgame}} are a highly popular type of game which feature a virtual world that allows players to freely modify the game world.
The canonical example of an \mve is Minecraft, which is the best-selling game of all time and supports over 140~million monthly active players~\cite{MinecraftOfficiallyCrosses2021}.
Although \mves are highly popular, their scalability in practice is very limited~\cite{DBLP:conf/wosp/SarDI19}, with commercial services partitioning players into small, isolated game instances and recommending as few as 10~(+1) maximum concurrent players per instance~\cite{MinecraftRealms2022Jan}.
Existing techniques for scaling online games can help increase the number of players for a variety of game types,
but do not address fundamental scalability challenges in \mves.
In contrast, in this work, we posit that \mve scalability can be addressed by leveraging serverless computing, and in particular, fine-grained deployment of applications as \faas.

\begin{figure}[t]
    \centering
    \includegraphics[width=\linewidth]{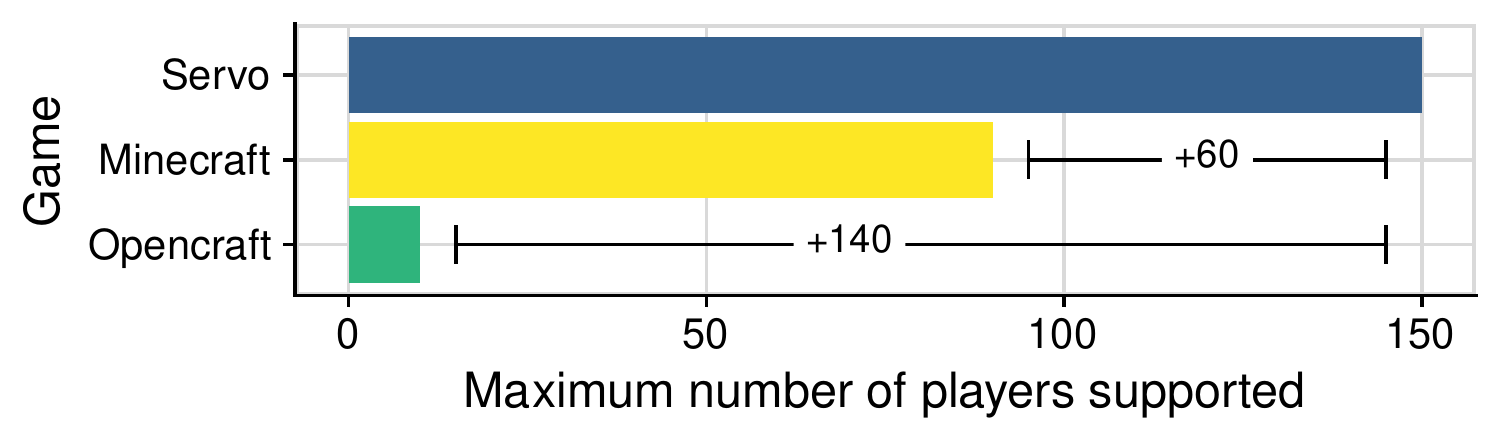}
    \caption{The maximum number of supported players in \sys.}
    \label{fig:teaser}
\end{figure}

Online games have existed since at least the PLATO\,III distributed system~\cite[L.4,195]{book:system:Dear17PLATO} from the 1970s.
Today, gaming is a lucrative industry that reaches a yearly revenue of \$175~billion~\cite{newzoo2021}.
An important part of the online gaming ecosystem are \emph{\acrlong*{mcgame}s}, which are online games that feature a virtual environment that is modifiable, or even programmable, by players~\cite{DBLP:conf/hotcloud/DonkervlietTI20}.
Prominent examples of \mves are Minecraft, Roblox, and Second Life.

Fundamental scalability challenges
in \mves include numerous elements that do not appear in other types of games:
its computationally-intensive virtual world containing in-game constructs that change state actively, player-created constructs that act as (repeated) programs,
and other aspects we detail in~\Cref{sec:model}.

A promising technology to address \mve scalability challenges is serverless computing~\cite{DBLP:journals/internet/EykTTVUI18, DBLP:journals/cacm/Schleier-SmithS21,DBLP:journals/cacm/CastroIMS19}.
Serverless computing is a cloud-computing technology and another step towards providing computational resources as a utility.
The canonical example of serverless computing is \faas,
which offers users serverless compute.
However, serverless services are known to exhibit high latency and latency variability,
which makes them difficult to use in real-time systems such as online games.

In this work, we posit that serverless computing, and in particular, fine-grained deployment of applications as \faas, can address \mve scalability when used in conjunction with latency-hiding mechanisms.
Starting from this design principle, we design, implement, and conduct real-world experiments with \sys, a novel \emph{serverless} backend architecture for scaling \mves.

Figure~\ref{fig:teaser} illustrates the real-world experimental results we obtain with a prototype of \sys, in comparison to Minecraft, the leading commercial \mve, and Opencraft, a state-of-the-art open-source \mve %
without a serverless architecture~\cite{DBLP:conf/icdcs/DonkervlietCI21}.
\sys scales to 150 players per instance, whereas Minecraft and Opencraft support up to 90 and 10 players,
which is an increase of 60 and 140 players respectively.

\textit{Existing approaches do not solve the challenge of scaling \mves{}.} %
\Cref{fig:teaser} also illustrates that \mve workloads can pose significant stress to the gaming platform for even relatively few players, through other in-game elements that create complex workloads dynamically (which we detail in~\Cref{sec:model:characteristics}).
These elements raise very diverse scalability challenges that we aim to address in this work.
Existing scalability techniques for online games~\cite{DBLP:journals/corr/abs-2102-09847, DBLP:journals/csur/LiuBC12, DBLP:journals/csur/LiuT14, DBLP:journals/tpds/GilmoreE12}, such as \emph{zoning} and \emph{replication}, can improve scalability for non-\mve{} games, but are not designed to take into account \mve{} scalability, which features complex in-game constructs that change state actively, player-created constructs that act as repeated programs, and other aspects we detail in~\Cref{sec:model}.
As we explain in~\Cref{sec:model:existing-approaches}, zoning leads to frequent communication and coordination between game-servers, and replication even multiplies the workload caused by active in-game components.

\textit{Scaling Minecraft-like \mves{} can have %
    high societal impact.}
Through their modifiable environments, \mves provide important benefits to society, providing entertainment, education,
activism, and social interaction to an audience of more than 100~million people.
With its large numbers of players and \emph{mods}, \mves provide world-wide entertainment,
but they are also used for other beneficial societal needs.
For example, \mves are used for education in academia~\cite{DBLP:conf/hci/WorsleyTMZJ21,DBLP:conf/chi/SlovakSTF18,DBLP:conf/acmidc/ZhuH17} and in industry with Microsoft's Minecraft Education
Edition~\cite{MinecraftOfficialSite}.
\mves are further used for social activism, with efforts including stopping illegal logging in Europe~\cite{Natividad2018Jan},
and promoting press freedom around the globe~\cite{UncensoredLibraryReporters}.
Minecraft is not alone as a popular \mve{}. In fact, tens of thousands of other \mves and \mve modifications (e.g., \emph{mods}) exist~\cite{mvesOnSteam, mcmods}, many with large followings of millions of active players.
Following this trend, Meta has already invested \$36~billion in developing an immersive, scalable \mve{} with advanced virtual-reality equipment for player-feedback~\cite{Meta2022}---a \textit{metaverse}~\cite{book:metaverse:ShowCrash92}.

\textit{Addressing the scalability challenges of \mves{} and focusing on Minecraft-like games, we design, prototype, and evaluate \sys}, %
a scalable, serverless, \mve system with fine-grained workload balancing.
Our main \textbf{C}ontributions are four-fold:
\begin{enumerate}[label=\textbf{C\arabic*}]
    \item We propose \mves as a promising but challenging use case for serverless, and particularly, \faas computing. We describe this use case in Section~\ref{sec:model}.
    \item We design \sys, a serverless backend architecture for \mves~(in Section~\ref{sec:design}).
          Ours is the first real-time online-multiplayer game using serverless design principles to improve the scalability of its virtual world. Although serverless simplifies building highly scalable and elastic applications, its programming model is
          not designed for online games, which operate under strict \qos requirements and need to continuously maintain
          consistency.

    \item We conduct real-world experiments evaluating the performance of our \sys prototype~(in Section~\ref{sec:experiments}). There is no standardized evaluation method for distributed systems, and
          benchmarking serverless systems is an active area of research~\cite{DBLP:conf/wosp/EykSEAI20,DBLP:conf/cloud/YuLDXZLYQ020}.
          We design comprehensive real-world experiments that can help compare serverless and non-serverless \mves{}, covering player-intensive workloads, workloads with complex in-game constructs, and other Minecraft-like workloads.
          From our key experiment on computational offloading~(\Cref{sec:experiments:simulated-construct-scalability}), we find that \sys can support up to 120~players during workloads with 200 in-game constructs, compared to 0~players(!) for Opencraft and Minecraft.
          For other workloads, \sys can
          increase the number of supported players under favorable conditions by 40~(+31\%) to 140~(+1400\%) players.
          These values summarize a complex suite of experiments, and not the exemplary result in \Cref{fig:teaser}.

    \item We implement a prototype on top of Opencraft, an open-source \mve and research platform. Our prototype runs on two popular commercial cloud platforms: \aws and Microsoft~Azure.
          We release its source code on Github: \ocgithub.
\end{enumerate}

\begin{figure*}[t]
    \centering
    \includegraphics[width=\linewidth]{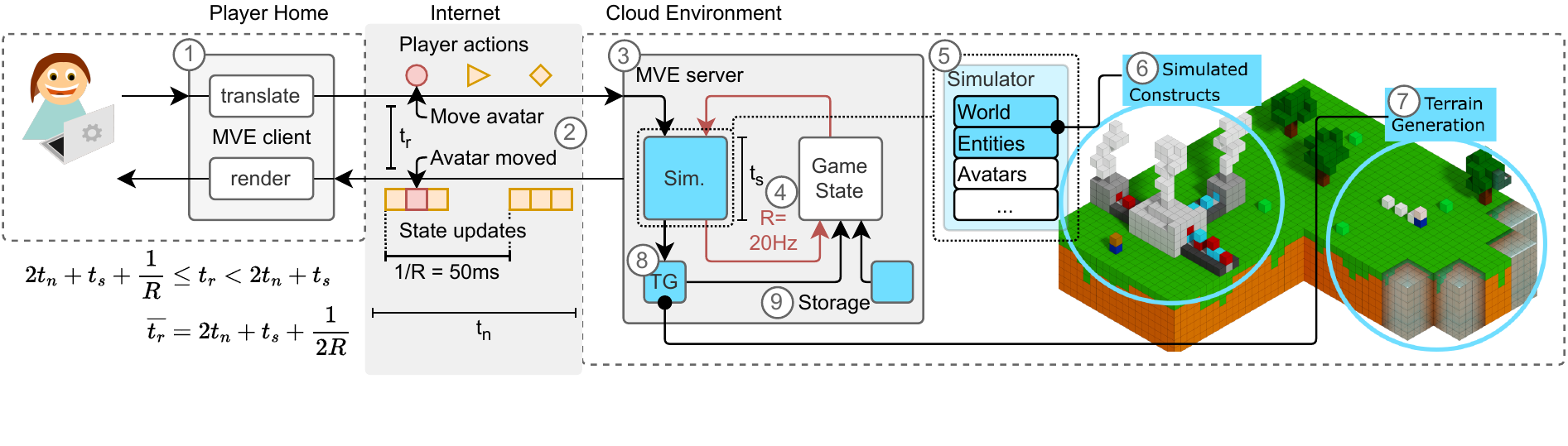}
    \vspace*{-1cm}
    \caption{An operational model of \mves. Elements highlighted in blue indicate the components most relevant to this work. Whiskers indicate latency values. Components \modelref{6}~and \modelref{7}~are examples of MVE workloads absent in traditional online games.}
    \label{fig:model:mve-workloads}
\end{figure*}

\glsresetall

\section{Modifiable Virtual Environments Use Case}
\label{sec:model}

We argue in this section that \mves are an important but challenging use case for serverless computing.
To this end, we propose and describe an operational model for \mves,
describe the performance characteristics of serverless computing,
and describe the benefits and challenges of using serverless computing for online games.

\subsection{The Modifiable Virtual Environment}
\label{sec:model:characteristics}

\mves are online real-time interactive systems, or games, which allow players to explore virtual worlds.
A main requirement of \mves is to provide stable and good performance to players.
To provide insight into how \mves meet this requirement, we propose and describe here an operational model of \mves, visible in \Cref{fig:model:mve-workloads}.

Users, or players, run an \emph{\mve client}~(\modelref{1}) on their personal device, which connects to an \emph{\mve server} over the Internet.
Players can only see and interact with other players connected to the same server.
The main responsibilities of the client are to translate and relay user inputs to the server,
and to show the user a visual representation of the virtual world by rendering state updates received from the server, typically at a fixed rate~(e.g., 60\,Hz).

Key to good performance is the game's response time ($t_r$),
which is the time between a player performing an action on their client,
and the result of this action becoming visible to all players connected to the same server.
This metric is shown visually in Figure~\ref{fig:model:mve-workloads} using a ``Move avatar'' command as an example~(\modelref{2}).
Although \emph{click-to-photon latency}, the time between a player clicking a button and the result appearing on the screen, is significantly affected by local and remote latency~\cite{DBLP:conf/mmsys/RaaenP15, DBLP:conf/chi/IvkovicSGS15},
we focus in this work on the distributed systems aspects,
modeling response time as dependent on the network latency ($t_n$) and the performance of the server ($t_s$).

Servers~(\modelref{3}) are typically deployed in a cloud datacenter.
Their main responsibility is to maintain a view of the global game state (for one instance of the virtual world) and simulate its state changes.
The game simulates an immersive virtual world using its \emph{simulator},
which takes as inputs player actions and the current game state,
and produces state changes which are sent to players and stored internally to be used in the next iteration of the simulator.

The performance of the server depends on two main factors~(\modelref{4}).
First, the time it takes for the simulator~(\modelref{5}) to complete one full iteration ($t_s$).
Second, the simulation rate~($R$), i.e., the number of iterations per second.
The simulation rate is typically fixed and varies per game type.
For first-person perspective games, $R$ is typically between 20 and 120, with Minecraft using an $R$ of 20.
Lowering the simulation rate increases the time budget for individual iterations,
but also increases response time.

Providing good performance to users requires maintaining good response time,
which requires maintaining the simulation rate, $R$, under varying workloads.
This is challenging because online games, and especially \mves, have highly dynamic workloads.
These workloads exert monthly, weekly, and diurnal patterns,
but also fluctuate moment-to-moment based on player behavior and environment simulation.
For game service providers to offer these services cost-efficiently,
this requires scalable and elastic resource management.
Analyses of the temporal workload dynamics of online games are available in existing work~\cite{DBLP:journals/tpds/NaeIP11}.
In this work, we focus on \mves.
Their modifiable, programmable, and procedurally generated terrain creates additional dynamic workloads for the server, which we describe below.

First, the game must process \emph{simulated constructs~(SCs,~\modelref{6})}.
Players can modify the terrain by placing and removing blocks of varying types.
Some block types, such as batteries and lamps, have internal state
and can interact with other nearby blocks.
For example, placing a battery next to a lamp may temporarily turn the lamp on.
By connecting a collection of stateful blocks, players can \emph{program} the \mve's terrain.
Throughout this article, we refer to such collection of stateful blocks as \emph{simulated constructs} or \emph{SCs}.
Players can construct SCs of arbitrary size, within the limits imposed by the game,
ranging from simple digital circuits to programmable digital computers,
creating additional work for server instances,
which must keep track of block states and their interactions with other blocks.

Second, procedural content generation~(PCG,~\modelref{7}) is an important task in \mves
because players explore an infinite world that is generated on demand.
When players move their avatar through the virtual world, the game must generate new parts of the world (i.e., terrain)
with low latency, before the terrain data must be forwarded to the client.

Once a part of the terrain is generated, it is managed and persisted by the game server.
Due to the world's virtually infinite size, existing terrain must be loaded to and from persistent storage dynamically, based on avatar locations.
Simulated constructs are only updated when their terrain is loaded in memory.
When players leave the area, the terrain state is persisted~(\modelref{9}) and its simulation is halted.
When players approach areas that are not in memory, the \mve must retrieve these areas from storage,
start updating any embedded simulated constructs, and forward the terrain data to clients. %
These actions must be completed before the terrain comes into the player's view,
and can increase the \mve's bandwidth usage by an order of magnitude~\cite{DBLP:conf/wosp/SarDI19}.

Although the generation and storage of terrain are not part of the game's simulator,
we identify two ways in which these elements can affect the game's performance.
First, a lack of performance isolation can lead to interference,
reducing the simulation rate when terrain generation is triggered.
Second, even when the game meets its intended simulation rate,
the separate generation component can fail to generate content at a sufficiently high rate,
preventing players from moving to new areas and making the game unplayable.

To manage system complexity, our model makes two simplifying assumptions.
First, it assumes network latency~($t_n$) is symmetrical.
Second, it assumes that state updates are only sent after the simulator has completed one full iteration.
However, implementations can interleave the sending of state updates and simulating state changes.

\subsection{\mve Architecture and Existing Scalability Approaches}
\label{sec:model:archi}
\label{sec:model:existing-approaches}

\mves typically follow a traditional client-server architecture, where players run client software on their local
device, which connects to a remotely-deployed server instance.
Each server instance provides access to an isolated and independent world instance, which means players can only
interact when their clients are connected to the same server.
In a traditional \mve, the scene depicted in Figure~\ref{fig:model:mve-workloads} is deployed on a single server and
several clients.

Although \mve instances use single servers, there are multi-server architectures for non-modifiable virtual worlds.
In this section, we describe two important classes of such cloud-based multi-server architectures: \emph{zoning} and \emph{replication}.
For an extensive overview of existing architectures for \mmos, see the following surveys~\cite{DBLP:journals/corr/abs-2102-09847, DBLP:journals/csur/LiuBC12, DBLP:journals/csur/LiuT14, DBLP:journals/ieeemm/MacedoniaZ97, DBLP:journals/tpds/GilmoreE12}.

Zoning is a technique to distribute the workload of an \mmo over multiple servers.
With zoning, the virtual world is partitioned into multiple \emph{zones}. Each zone is assigned to a server, making the
server responsible for simulating the avatars in that zone.
When an avatar crosses a border from a zone A to a zone B, A's server hands over the simulation of that avatar to B's
server.
When a player interacts with a player in another zone (e.g., when two avatars are on either side of a border), this
interaction results in server-to-server communication.
Although this works well for the mostly static environments used in non-\mve{} games,
\mves simulate also the environment itself, which, when partitioned across zones, requires frequent communication and coordination between
servers.

Replication is an alternative technique where the avatars are partitioned into multiple groups. Each group of avatars is
assigned to a server, making that server responsible for simulating the avatars in that group.
In contrast to zoning, replication allows the simulation of an avatar to remain with one server during the player's
session. However, because replication does not guarantee that avatars that are close to each other are handled by the
same server, player interaction can frequently result in server-to-server communication.
For \mves{}, replication
does not address, and even duplicates, the workload caused by the virtual environment itself, which is problematic given its
unbounded nature and computational intensity.

\subsection{Serverless Computing}
\label{model:faas}
\label{model:serverless}

Serverless computing is an emerging computing service and a next step towards making digital services as simple to use as traditional utilities.
Serverless computing promises effortless and cost-efficient scalability, but its characteristics make it challenging to adopt for certain application domains.

Although serverless computing does not have a generally-accepted precise definition,
we follow in this work the definition from~\cite{kounev_et_al:DagRep.11.4.34:ServerlessNotion}.
Following this definition, serverless computing:
\begin{inparaenum}[(1)]
    \item Moves the responsibility of \rms from the end user to the cloud operator. Users merely provide application logic, and trust the cloud operator to scale the amount of resources to the current workload.
    \item Employs fine-grained utilization-based billing. Users do not pay setup cost, or cost for idle resources, but only pay for resources used, with fine granularity (e.g., milliseconds).
\end{inparaenum}

Commercial clouds offer a plethora of serverless services.
Well-known examples of these services include \faas (e.g., AWS~Lambda and Azure~Functions) and serverless storage (e.g., AWS~S3 and Azure Blob Storage).
With \faas, users upload code to their cloud provider of choice and configure one or several events that trigger the execution of this code.
When such an event occurs, the cloud provider schedules the execution of the user-provided code.
This model has several appealing properties.
First, it allows for simple but powerful scalability because every request is handled by its own instance.
Second, the user only pays for the amount of time and resources consumed while executing their provided code, which means the user does not pay for setup costs nor gets charged when there are no requests.

Serverless storage allows users to read and write (binary) data to persistent storage in the cloud.
Depending on the service, users are charged for the amount of data stored, the number of read and write operations, the number of bytes transferred over the network, or a combination of these metrics.
In return, the cloud provider performs all necessary \rms, leaving the user with a simple read-write API.
Commercial clouds offer a range of serverless storage services that make different performance-cost trade-offs.
For example, services catering to long-term data archiving are generally cheaper to use, but have significantly higher read and write latencies, sometimes in the order of days.

\subsection{Leveraging Serverless for MVEs}
\label{sec:model:combine}
\label{sec:model:usecase}

In this section, we describe the main benefits and challenges of using serverless computing for online real-time games generally, and \mves specifically.

\subsubsection*{Benefits}

First, games can benefit from the simple elasticity serverless computing provides because gaming workloads change significantly over time~\cite{DBLP:conf/sc/NaeIPPEF08}.
Furthermore, \mve workloads can increase or decrease significantly in the order of seconds, depending on player actions.
For example, if a player builds and enables simulated constructs, a single player can exceed the resource requirements of a game instance~(see~\Cref{sec:experiments}).
Second, fine-grained billing allows this elasticity to be cost-efficient and allows smaller development teams to participate in the market without large up-front investments.

\subsubsection*{Challenges}
First, whereas games typically have stringent latency requirements in the order of tens of milliseconds,
serverless services commonly exhibit high latency and significant performance variability, for example due to \emph{cold starts}.
Not meeting the game's latency requirements can result in significant (visual) glitches for the player that can
cause them to quit the game, causing the game developer to lose users and revenue.

\begin{figure}[t]
    \centering
    \includegraphics[width=\linewidth]{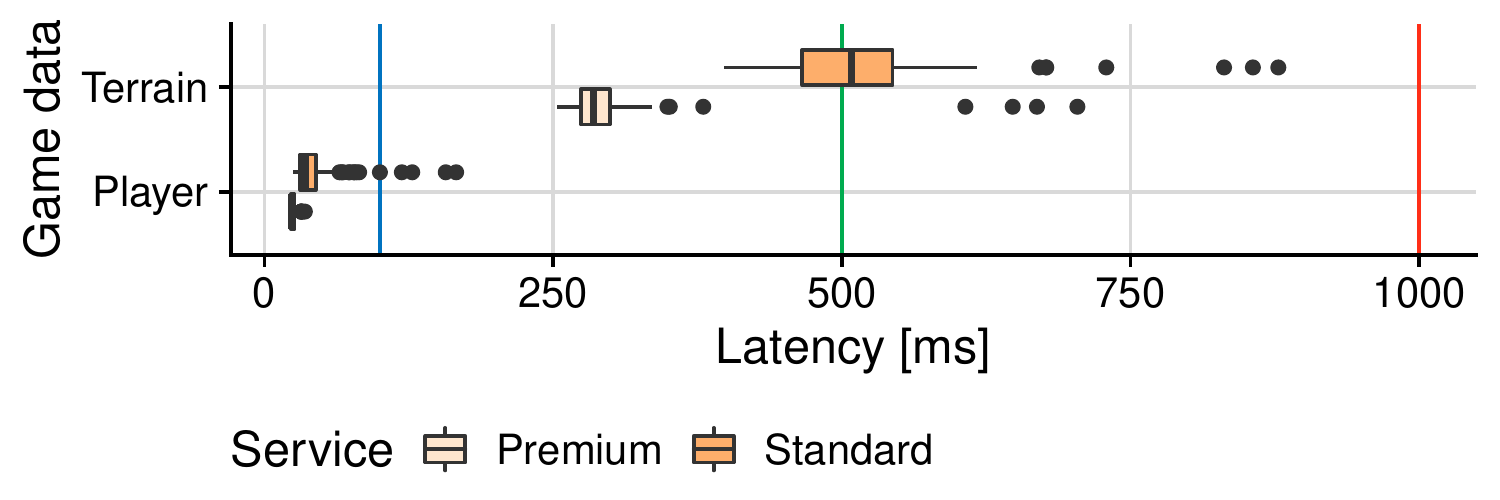}
    \caption{Download latency (variability) from Azure Blob Storage for two types of game data.Vertical bars indicate approximate network latency thresholds for FPS~(blue, left), RPG~(green, middle) and RTS~(red, right) games~\cite{DBLP:journals/cacm/ClaypoolC06}.}
    \label{fig:res:storage-azure-latency}
\end{figure}

An example of this phenomenon is shown in Figure~\ref{fig:res:storage-azure-latency}.
The figure shows the end-to-end latency of retrieving terrain-data and player-data from Azure Blob Storage.
The former operation is performed when players move their avatars towards a previously unloaded area of the virtual world,
which occurs frequently during play,
whereas the latter is performed every time a player connects to a game instance, and their associated data is retrieved from
persistent storage.
For both storage plans, the latency variability is significant, as indicated by the width and length of the box an whiskers, and the value of the outliers.
The vertical lines show approximate values for the maximum acceptable network latency for varying game genres~\cite{DBLP:journals/cacm/ClaypoolC06}.
Games with an omnipresent model (e.g., RTS games, in red),
third-person games (e.g., RPGs, in green),
and first-person games (e.g., FPSs, in blue).
Most \mves, including Minecraft, are first-person games,
putting them in the games with the most restrictive network latency requirement~(100\,ms).
This indicates that a naive application of serverless technology can easily break its performance requirements.

Second, \faas services typically impose strict limits on the amount of time a function is allowed to run, whereas online games typically consist of long-running player sessions that can last several hours.
Third, and finally, \faas services typically do not allow external services to initiate communication to a running function instance, whereas games and other interactive systems require continuous communication between the system components to maintain a consistent state.

\section{System Design}
\label{sec:design}

In this section we discuss our design for \sys,
a serverless backend architecture for \mves which uses a collection of serverless techniques to provide fine-grained virtual-world scalability.
We describe the requirements for such a system and formulate a novel design.

\subsection{System Requirements}
\label{sec:design:requirements}

\begin{enumerate}[label=\textbf{R\arabic*}]
    \item\label{req:independent-workload} \textbf{Independently scale \mve server components.}
          Although replicating monolithic \mve instances is the simplest approach to scale out, it is not efficient.
          Because players increase the workload on specific components of the \mve~(e.g., terrain simulation, terrain generation), scaling these components independently limits wasted resources.
    \item\label{req:qos} \textbf{Simulation latency should not exceed 50\,ms.}
          To provide an immersive experience to players, the game should maintain a constant simulation rate, $R=20$\,Hz. This implies the simulation latency should not exceed $1/R=50$\,ms~(see Section~\ref{sec:model}).
    \item\label{req:developer-operations} \textbf{Limit additional resource management complexity for game operators.} An important benefit of the client-server architecture discussed in Section~\ref{fig:model:mve-workloads} is its simplicity: a game operator runs one server instance for every virtual world instance. A serverless game design should limit the additional number of steps game operators must perform.
    \item\label{req:user-transparent} \textbf{System is fully transparent to users.} To improve usability and potential for community adoption, using the system should require zero effort from, and be fully transparent to, the users (i.e., players). Specifically, this means players should not need to run additional software or perform additional steps to benefit from the system.
\end{enumerate}

\subsection{Design Overview}
\label{sec:design:overview}

\begin{figure}
    \centering
    \includegraphics[width=\linewidth]{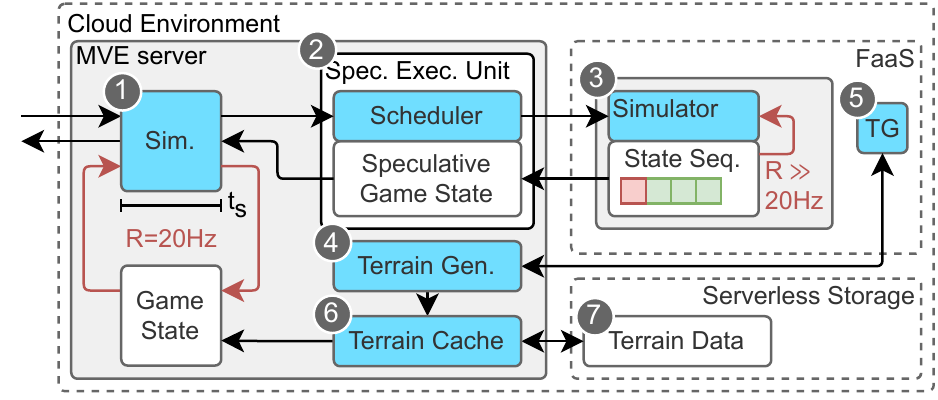}
    \caption[\sys design overview]{\sys design overview. Simulated constructs~(SC) and components \designref{1}, \designref{2}, etc. are discussed in the main text.}
    \label{fig:sys:design}
    \label{fig:design:overview}
\end{figure}

This section describes the design of \sys,
a serverless backend architecture for \mves that provides fine-grained horizontal scalability while maintaining low operational effort from game developers.
Figure~\ref{fig:sys:design} shows our design, with novel aspects indicated in blue.

\sys is a backend system that modifies and interacts with server instances.
It does not interact with the game client, nor change the protocol used between the client and server~(addresses~\ref{req:user-transparent}).
Instead, \sys modifies existing components in the server to leverage serverless services.
To save resources and maintain an update rate of 20\,Hz under high workload,
\sys uses computational offloading for simulated constructs, a computationally intensive part of the game loop~(first part of Requirement~\ref{req:independent-workload}).

\emph{Novel in this work}, \sys leverages serverless technologies such as Function-as-a-Service~(FaaS) and managed storage
to provide fine-grained scalability for the key \mve components described in~\Cref{sec:model:characteristics}.
We provide first an overview of \sys's novel elements,
and proceed by describing each of these elements in detail in the remainder of this section.

To provide good performance for terrain simulation, \sys modifies the \mve simulator~(\designref{1}) to use a \emph{speculative execution unit}~(\designref{2}) to simulate SCs.
The speculative execution unit is responsible for managing speculative executions of each simulated construct, and providing the simulator with the speculative states that turn out to be correct.
The execution unit offloads the SCs to serverless functions~(\designref{3}, partially addresses Requirement~\ref{req:developer-operations}).
\emph{Unlike the game loop, the function is not required to simulate changes at a fixed rate (e.g., 20\,Hz),
    allowing it to compute state changes speculatively.}
Each function invocation simulates a single simulated construct, computes its state changes for multiple simulation steps, and sends these state changes back to the execution unit, allowing the simulator to merge the state changes with those computed in the game loop~(partially addresses~\ref{req:qos}).

To independently scale terrain generation,
\sys moves the responsibility of terrain generation from the (monolithic) server to an independent serverless function,
reducing the server's computational load~(completes~\ref{req:independent-workload}).
The \emph{terrain generator}~(\designref{4}) is responsible for invoking terrain generation tasks based on the current
location of each avatar.
Each generation task~(\designref{5}) corresponds to a serverless function invocation, and generates a fixed-sized area of the terrain.

Finally, \sys moves the responsibility of managing persistent game state from the game operator to the cloud provider,
storing game state in cloud-based managed storage~(\designref{7}, \ref{req:developer-operations}).
\mve persistent storage includes player-, meta-, and terrain-data.
We focus in our design on the latter because it is the most data-intensive and is accessed most frequently.
To maintain good \qos, \sys uses a server-local cache~(\designref{6}) with a simple pre-fetching policy
to hide latency and performance variability from the managed storage solution~(\ref{req:qos}).

\subsection{Replicated Speculative Execution for Simulated Constructs}
\label{sec:design:sc}

\mves allow players to build \emph{simulated constructs}.
As described in \Cref{sec:model}, simulated constructs are part of the virtual world's terrain, allow players to program the virtual world,
and generate additional computational load for the \mve instance.
This section describes how \sys uses computational offloading and speculative execution to scale simulated constructs.

\sys scales horizontally by offloading the computation of individual simulated constructs to serverless functions.
Whenever a player builds and activates a simulated construct, \sys invokes a function and passes the simulated construct's current state
and the number of steps to simulate.
Although this approach scales well with the players' ability to build an endless number of simulated constructs,
it raises an important design challenge:
Because simulated constructs are part of the terrain, their update rate must match that of the virtual world~(i.e., 20\,Hz),
and their state changes must become visible to players with low latency~(i.e., after a single tick, or 50\,ms).

\begin{figure}[t]
    \centering
    \includegraphics[width=\linewidth]{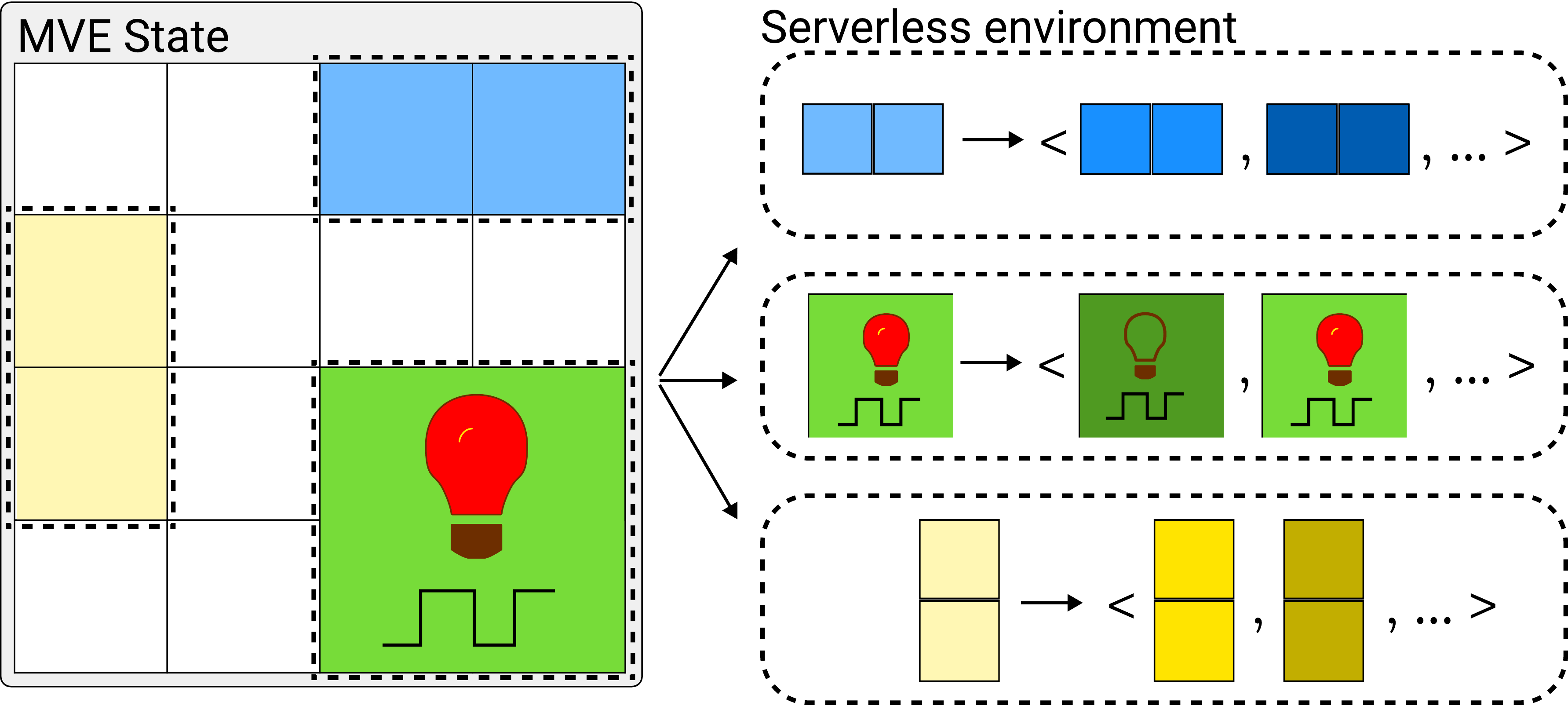}
    \caption{A visual representation of speculative execution in \sys. Colored areas are simulated constructs. White areas are static parts of the virtual world.}
    \label{fig:design:simulated-construct-offloading}
    \vcutS
\end{figure}

\sys addresses this challenge through replicated, speculative execution.
Figure~\ref{fig:design:simulated-construct-offloading} presents an overview of this approach.
The left-hand side of the figure depicts a simplified top-down view of a voxel-based 3D virtual world.
The colored areas are simulated constructs: simulated elements that change state.
An example of a simulated construct is a circuit that powers a lamp, shown in green and red.
\sys isolates the state of these elements and forwards each to a serverless function for speculative execution.
If no speculative state results are available on the game server on time, or the speculative state is incorrect,
\sys falls back to server-side simulation.

Speculative results can be incorrect because, although SCs operate independently of other simulated constructs and players,
players have the possibility to interact with the simulated construct, for example by modifying the underlying terrain.
Because such changes cannot be part of the original request to the remote function,
its results will likely be inconsistent with the new correct state.
Addressing this challenge, we include in the request a logical timestamp indicating when a player last modified the simulated construct.
The function sends this same timestamp in its reply, allowing the server to disregard the reply if its result is outdated.

This approach reverses the common application of speculative execution and rollback mechanisms:
instead of speculatively applying updates locally and undoing those changes if they turn out to be incorrect,
\sys performs speculative execution remotely and only applies changes when they are correct.
This approach is feasible due to the game's fixed update rate,
which artificially limits their processing speed.

\begin{figure}[t]
    \centering
    \includegraphics[width=\linewidth]{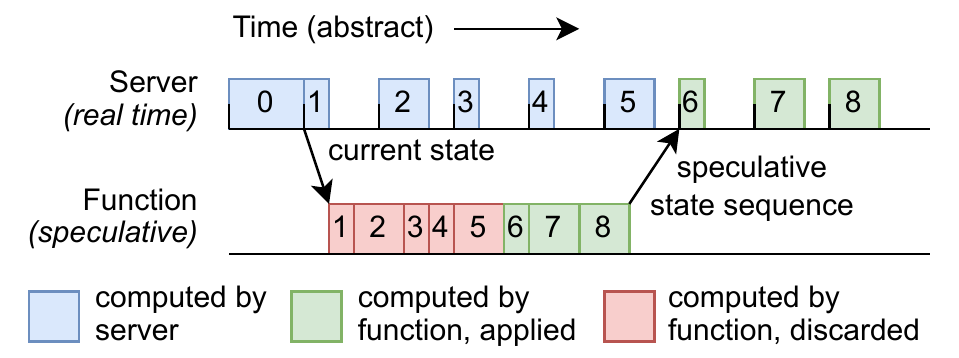}
    \caption{Example of speculative execution for simulated constructs. \sys simulates the construct locally, until it receives the results from the remote function.}
    \label{fig:simulated-construct-compute-ahead}
    \vcutS{}
\end{figure}

Figure~\ref{fig:simulated-construct-compute-ahead} shows a visual representation of this approach.
When a simulated construct is started, \sys starts simulating simultaneously on the server and in a remote function.
Starting the simulation on the server hides the (cold-start) latency from the remote function~(completes~\ref{req:qos}),
However, performing all the work on the server would defeat the purpose of offloading the computation.
Instead, we observe that, unlike the server, the function does not need to simulate at a fixed rate,
and therefore can (speculatively) work ahead.
This allows the function to catch up with, and get ahead of, the game server, and return a collection of future states.
Upon receiving a reply from the worker, \sys stops its local simulation and
instead applies state updates received from the remote function.

The number of steps that need to be computed locally determines the \emph{efficiency} of the speculative execution.
For the example shown in Figure~\ref{fig:simulated-construct-compute-ahead}, 5 out of 8 steps are computed both locally and in the offloaded function, resulting in an efficiency of $\frac{8-5}{8}\approx0.38$.
Although the first speculative execution is likely to reach low efficiency, the efficiency is likely to increase significantly for all following invocations by invoking further speculative execution several steps before the results are needed.
Throughout the rest of the article, we refer to this number of steps as the \emph{tick lead}.
Returning to the example in Figure~\ref{fig:simulated-construct-compute-ahead}, the second speculative execution can start at time 6, but simulate starting at state 8, resulting in a tick lead of 8-6=2.

\begin{techrep}
    \subsubsection{Cost Optimization: Loop Detection}
    \label{sec:impl:sc-loop-detection}

    Horizontally scaling the virtual world can increase cost for the game operator.
    We discuss here a cost optimization for simulated constructs that loop through a list of states indefinitely,
    such as logical clocks and certain types of resource farms.
    By only simulating the looping behavior once, and replying the state changes afterwards,
    the \mve can reduce resource usage and cost.

    To this end, our remote simulation function records the result of every simulation step (i.e., tick) using a hashing function.
    If the simulated construct repeats earlier seen states, the function truncates the resulting states to one iteration of the loop,
    and adds an index pointing to the construct's current state.
    This allows \sys to loop through the constructs state changes repeatedly, without simulating (locally or remotely) its
    behavior.
\end{techrep}

\newcommand{\sw}[1]{$\mathds{S}$#1}
\newcommand{\aw}{$\mathds{A}$}
\newcommand{\rw}{$\mathds{R}$}
\newcommand{\highlightcell}{}%
\begin{table*}[t]
    \centering
    \caption{Overview of Experiments. Experiments focus on Simulated Constructs~(SC), Terrain Generation~(TG), and Remote Storage~(RS). Components either run locally (L), use serverless computing (S), or combine the two (L+S).
        For a detailed description of parameters, workload, environment, and explanations on the notation, see~\Cref{sec:exp:setup}.}
    \label{tab:experiments}
    \begin{tabulary}{\linewidth}{lLlllrRlllllrr}
        \toprule
        & Focus                  & \multicolumn{3}{c}{Parameters} &  &                   \multicolumn{3}{c}{Workload}                    &  & \multicolumn{2}{c}{Environment} &  Rep. & Dur. \\
        \cmidrule{3-5} \cmidrule{7-9} \cmidrule{11-12} &                        & SC                 & TG  & RS  &  & Players            & Behavior                           & World   &  & Server & Services                   &       &  [m] \\ \midrule
        \Cref{sec:experiments:simulated-construct-scalability} & SC: System scalability & \highlightcell L+S & L   & L   &  &  10-200 & \aw                                & flat    &  & \das   & AWS                    &     1 &    10 \\
        \Cref{sec:experiments:sc-efficient} & SC: Latency hiding            & L+S                  & L   & L   &  & 1 & -                             & flat    &  & \das   & AWS                    &     1 &    5 \\
        \Cref{sec:experiments:terrain-gen-aws} & TG: QoS            & -                  & S   & L   &  & 5                  & \sw{inc}                             & default &  & \das   & AWS                    &     1 &    5 \\
        \Cref{sec:experiments:serverless-improves-scalability} & TG: System scalability     & -                  & L+S & L+S &  & 100 & \highlightcell \sw{3}, \sw{8}, \rw & default &  & Azure  & Azure                  & 1, 20 &   10 \\
        \Cref{sec:experiments:storage-cache-policy} & RS: Perf. variability  & -                  & -   & S   &  & 8                  & \sw{3}                             & default &  & Azure  & Azure                  &     1 &   10 \\
        \Cref{sec:experiments:offloading-sc-size-performance} & SC: Performance & S & - & - & & 1 & - & flat & & \das & AWS & - & - \\
        \bottomrule
    \end{tabulary}
    \vcutM{}
\end{table*}

\subsection{Leveraging FaaS for On-Demand Content Generation}

Players exploring the virtual world, or teleporting their avatar to new locations, cause computationally intensive workloads and contention for resources with the game's simulator, which must maintain stable performance.
Here we briefly describe how \sys horizontally scales the \mve's content generation process by running content generation tasks in remote functions.

As described in~\Cref{sec:model}, \mve's generate new areas of the terrain on-demand, when players approach these areas.
\sys separates the virtual world and content generation responsibilities and moves the content generation component to a remote function.
When players approach non-existent areas, the game invokes a remote function for each area that requires generation,
providing as parameters the seed for the pseudo-random number generator, and the coordinates of the area.
Although the generation of a single area is expected to complete in the order of seconds, all generation requests can be invoked concurrently, leveraging \faas's simple scalability.

\begin{table}[t]
    \centering
    \caption{Player actions in the random behavior~($\mathds{R}$). See \Cref{sec:exp:setup} for a detailed description.}
    \label{tab:player-behavior}
    \begin{tabulary}{\linewidth}{rL}
        \toprule
        Probability & Action  \\ \midrule
        40\% & \textbf{Move} to a random destination at 1 to 8 blocks per second.  \\
        30\% & \textbf{Break or place} a nearby block. \\
        20\% & \textbf{Stand} still. \\
        5\% & \textbf{Send a message} to all other players. \\
        5\% & \textbf{Set inventory} to a random item. \\
        \bottomrule
    \end{tabulary}
    \vcutM{}
\end{table}

\subsection{Remote State Storage}
\label{sec:design:storage}

Whereas \mves typically save state to local disk,
\sys uses serverless storage to reduce the operational effort for the game developer (completes~\ref{req:developer-operations}).
However, managed storage can exhibit significant performance variability,
which must be hidden from \mve players to provide good \qos.
Although \sys uses managed storage for player-, meta-, and terrain-data, we focus in this section on terrain-data,
because it is the most challenging:
terrain-data consists of relatively large files which are accessed frequently and with strict latency requirements.

\sys addresses the latency requirements of terrain storage by caching terrain data on local storage and pre-fetching data from remote storage.
When avatars change location, the storage service checks which files correspond to the avatar's current location and the areas in the player's view range,
and load these into memory from the local file system or remote storage.
Areas not within any player's view distance are removed from memory after a certain amount of time.
\sys caches these areas on the local file system, reducing the number of accesses to remote storage and saving bandwidth.
To improve the cache's hit rate and hide latency, \sys pre-fetches terrain data outside of, but close to, the player's view distance and
proactively loads these into memory.
In contrast to reads, writes to remote storage are performed periodically.

\section{Experiments}
\label{sec:experiments}

To evaluate our design, we implement \sys on top of Opencraft, an open-source MVE and research platform compatible with the original Minecraft network protocol~\cite{DBLP:conf/icdcs/DonkervlietCI21}.
We implement \sys's serverless components on the two most popular commercial cloud platforms: Amazon Web Services~(AWS) and Microsoft Azure, and conduct real-world experiments using our prototype implementation of \sys.
Table~\ref{tab:experiments} shows an overview of our experiments, from which we derive the following main findings:

\begin{enumerate}[label=\textbf{MF\arabic*}]
    \item Serverless offloading of simulated constructs improves game scalability. \sys supports up to 140~(+1400\%) additional players under favorable conditions, compared to state-of-the-art alternatives~(\Cref{sec:experiments:simulated-construct-scalability}).
    \item Speculative execution efficiently hides serverless offloading latency, with a median of 84\% of speculative states applied without invoking functions in advance, and 100\% when invoking functions 10 steps in advance~(\Cref{sec:experiments:sc-efficient}).
    \item Serverless content generation provides good \qos~(\Cref{sec:experiments:terrain-gen-aws}). \sys can maintain full view distance (128) whereas Opencraft drops below 16.
    \item Serverless content generation maintains good performance, increasing the number of supported players by 6 compared to Opencraft~(\Cref{sec:experiments:serverless-improves-scalability}).
    \item Caching significantly reduces performance variability when reading game data. \sys reduces the 99.9\textsuperscript{th} percentile latency for loading terrain data from serverless storage from 226\,ms to 34\,ms, enabling terrain loading within a single simulation step~(\Cref{sec:experiments:storage-cache-policy}).
    \item Simulated constructs of small and medium scale can be used for speculative execution effectively. At least 95\% of samples speculatively executing 100 steps for simulated constructs consisting of 252 and 484 blocks simulate at a rate of 488 and 105 updates per second,
          24.4 and 5.3 times faster than the simulation rate respectively~(\Cref{sec:experiments:offloading-sc-size-performance}).
\end{enumerate}

\subsection{Experimental Setup}
\label{sec:exp:setup}
\label{sec:experiments:setup}

\begin{figure*}
    \begin{subfigure}[t]{.24\textwidth}
        \includegraphics[width=\linewidth]{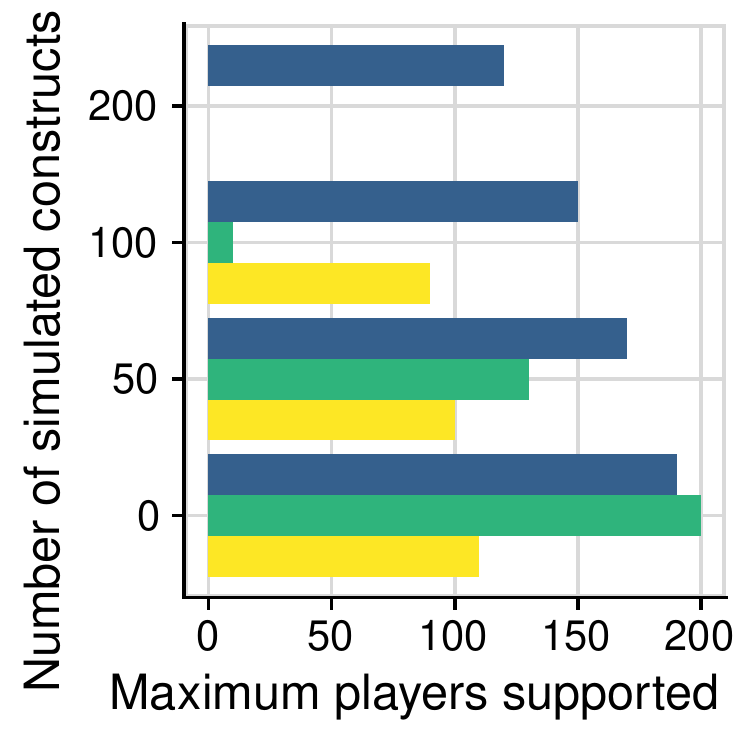}
        \caption{Max. players for increasing SCs. Higher is better.}
        \label{fig:res:sc-aws:scalability}
    \end{subfigure}
    \hfill
    \begin{subfigure}[t]{.72\textwidth}
        \includegraphics[width=\linewidth]{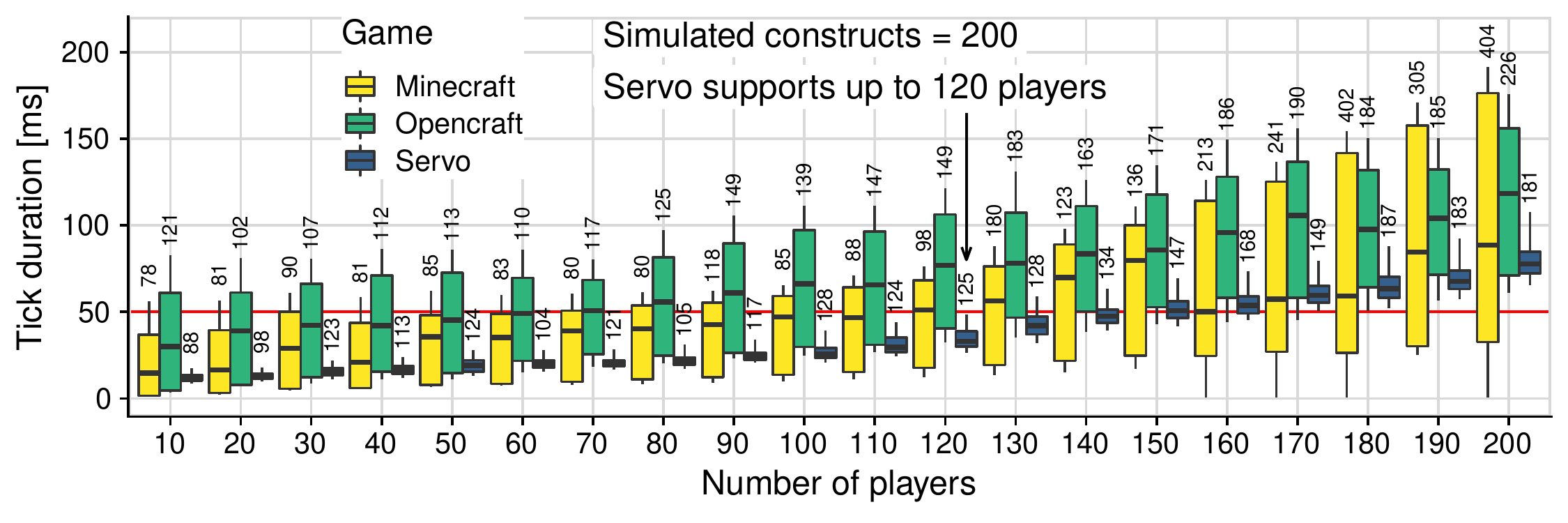}
        \caption{The tick duration distribution for a varying number of players and 200~SCs. Lower is better.}
        \label{fig:res:sc-aws:boxplot}
    \end{subfigure}
    \caption{Comparing the scalability of \sys with alternative systems. Scalability is expressed in maximum number of players while maintaining stable performance. Missing bars indicate the game could not support any players.}
    \label{fig:res:sc-aws}
    \vcutM{}
\end{figure*}

Table~\ref{tab:experiments} shows an overview of our experiments, and details their parameters, workloads, and experiment environment.
The experiments cover three major system parameters: \emph{simulated constructs~(SC in~\Cref{tab:experiments}, explained in~\Cref{sec:model:characteristics} and depicted in~\Cref{fig:model:mve-workloads}, component 6)}, \emph{terrain generation~(TG)}, and \emph{remote storage~(RS)}. Each of these components either runs locally~(L in~\Cref{tab:experiments}), uses serverless computing~(S), or a combination of the two~(L+S).

The experiment workloads are defined through three parameters.
First, \emph{players} indicates how many players connect to the game throughout the experiment.

Second, the \emph{behavior} indicates what players do once connected.
$\mathds{S}x$ indicates that players move away from the starting location (i.e., spawn location) at a speed of $x$ blocks per second, in a star-shaped pattern,
whereas $\mathds{S}$inc indicates players that increase their velocity over time.
These synthetic behaviors increase the amount of terrain covered by the players, and allows stress-testing the game's terrain generation.
$\mathds{R}$ indicates a randomized behavior, in which players perform actions that are common in \mves.
An overview of these actions is available in Table~\ref{tab:player-behavior}.
We use a randomized behavior because no \mve-specific player behavior models currently exist.
$\mathds{A}$ indicates a behavior where players exclusively take move actions, within a bounded area. This behavior limits the amount of terrain generation,
and is used to evaluate simulated construct~(SC) performance.

Third, the \emph{world} indicates the type of terrain (generation) used.
The \emph{default} world type uses procedural content generation and is commonly used by real players and features mountains, rivers, and other natural phenomena.
The \emph{flat} world type consists of an infinite plain in all directions. This terrain makes it easy to build, and players commonly use this world type to prototype simulated constructs.

We perform our experiments on the commercial cloud platforms \aws and Microsoft Azure,
and on \das, a local medium-sized supercomputer for research and education~\cite{DBLP:journals/computer/BalELNRSSW16}.
\emph{Server} indicates the location of the \mve server instance, and \emph{services} indicates the platform used for managed services (e.g., FaaS, managed storage).
Unless otherwise indicated, we compare the performance of \sys with Opencraft
Microsoft's official Minecraft~1.16 Java Edition.

\subsection{Offloading Simulated Constructs Improves \mve Scalability}
\label{sec:experiments:simulated-construct-scalability}

Our results show that:
(1) Using serverless offloading of simulated constructs, \sys can support up to 120~players under workloads where Opencraft and Minecraft support none~(0);
and~(2)~For other workloads tested, \sys can increase the number of supported
players under favorable conditions by 40~(+31\%) to 140~(+1400\%)
compared to Opencraft and Minecraft.

Figure~\ref{fig:res:sc-aws:scalability} shows the maximum number of supported players for varying workloads.
The vertical axis shows four workloads with an increasing number of simulated constructs (i.e,. increasing computational requirements).
The horizontal axis shows the maximum number of supported players,
which we define as the maximum number of players for which less than 5\% of tick duration samples exceeds 50\,ms.
The bars indicate different games.
\sys (blue) serverlessly offloads the simulation of the simulated constructs.
\sys is based on Opencraft (green), which computes simulated constructs locally.
Minecraft (yellow) is the official Minecraft server.

The result shows that \sys significantly improves scalability in nearly all cases,
and introduces only limited overhead.
The workload with 0 simulated constructs shows the baseline performance.
Minecraft supports 110 players in this scenario, whereas \sys and Opencraft support~190 and 200~players respectively.
For higher numbers of simulated constructs, the maximum number of supported players decreases for all games,
but \sys performs best under these conditions.

Opencraft and Minecraft support up to 10 and 90 simultaneous players respectively in worlds with 100 simulated constructs.
\sys supports 150 players, an increase of~140~(+1400\%) and~60~(+67\%) players respectively.
When using 200 SCs, \sys supports~120 players whereas Opencraft and Minecraft support zero~(0).
Even when connecting a single player to Opencraft or Minecraft, their 95\textsuperscript{th} percentile tick duration exceeds 50\,ms.

Figure~\ref{fig:res:sc-aws:boxplot} shows the detailed performance of all three games when using 200~SCs
for multiples of 10 connected players between 0 and 200 (corresponds to the top-row in Figure~\ref{fig:res:sc-aws:scalability}),
and allows an analysis of the observed scalability differences.
We further evaluate the scalability of the games for fewer than 10 players, but omit this data from the plot for improved readability.
The horizontal axis shows the number of connected players,
and the vertical axis shows a boxplot of the game's tick durations.
The whiskers extend to the 5\textsuperscript{th} and 95\textsuperscript{th} percentiles.
The maximum value is shown in the text above each box, and further outliers are omitted.
To maintain good and stable performance, the tick duration must be below 50\,ms.
The red line indicates the 50\,ms threshold.

We observe that for both Minecraft and Opencraft, a significant number of samples exceed 50\,ms,
and that this number increases with the number of connected players.
In contrast, \sys's tick duration is significantly lower, and shows a narrower distribution (box height).
\sys supports up to 120 players because fewer than 5\% of tick duration samples exceed the 50\,ms threshold up to this number of players.

We observe that \sys's tick duration is consistently close to the 25th percentile of Opencraft's tick duration.
Upon further analysis, we observe that both Opencraft's and Minecraft's tick duration distributions are bimodal,
and that this behavior is caused by their implementation, in which simulated constructs are simulated every other tick.
I.e., the bottom half of their distribution shows the performance when not updating the simulated constructs.

Overall, the results show that offloading significantly improves the performance of the \mve.

\begin{techrep}

\end{techrep}

\subsection{Speculative Execution Effectively Hides FaaS Latency}
\label{sec:experiments:sc-efficient}

\begin{figure}[t]
    \centering
    \begin{subfigure}[t]{0.49\linewidth}
        \includegraphics[width=\linewidth]{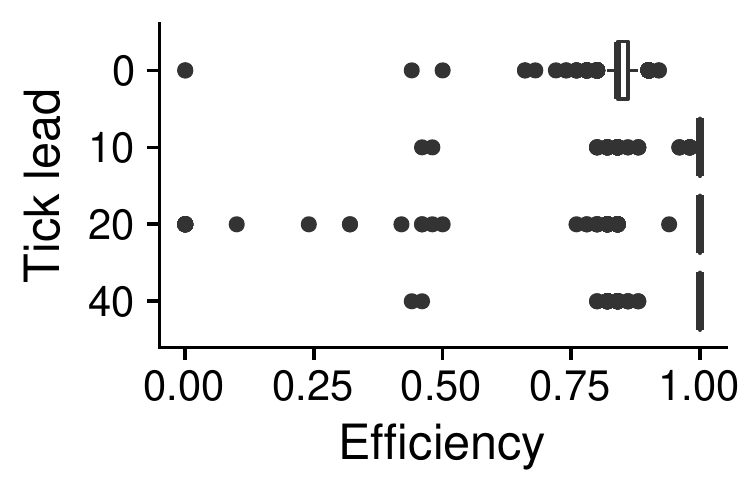}
    \end{subfigure}
    \begin{subfigure}[b]{0.49\linewidth}
        \includegraphics[width=\linewidth]{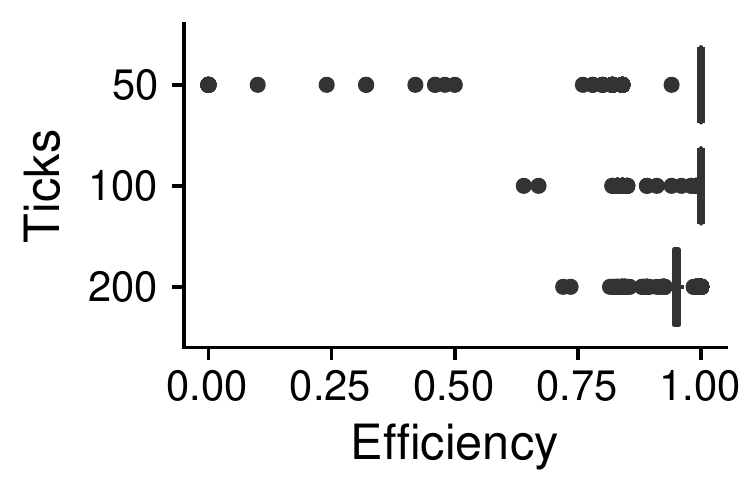}
    \end{subfigure}
    \caption{Efficiency of offloaded simulation for varying tick leads~(left), and simulation lengths~(right).}
    \label{fig:res:offloading-efficiency}
    \vcutM{}
\end{figure}

Our results show that \sys's latency hiding mechanism can effectively hide the latency from serverless functions.

Figure~\ref{fig:res:offloading-efficiency} shows this result.
Both plots show the \emph{efficiency}~(\Cref{sec:design:sc}) of the offloaded computation on the horizontal axis.
The efficiency is computed as the fraction of offloaded work that must still be performed by the game server because its result arrives too late at the server instance.
The left-hand plot shows the tick lead on the vertical axis, which shows how many ticks in advance of the computation result being required the function is invoked.

The results show that a tick lead of 0 leads to a median efficiency of 84\% (top row in the left-hand plot),
indicating that 16\% of the offloaded computation must still be performed locally to meet the game's latency requirements.
However, when the function is invoked in advance by 10, 20, or 40 ticks, median efficiency reaches 100\% (boxplots align vertically at 1.00).
Further analysis shows that for these three configurations, at least 99.1\% of invocations have an efficiency of 100\%.

\begin{figure}[t]
    \centering
    \begin{subfigure}[t]{0.49\linewidth}
        \includegraphics[width=\linewidth]{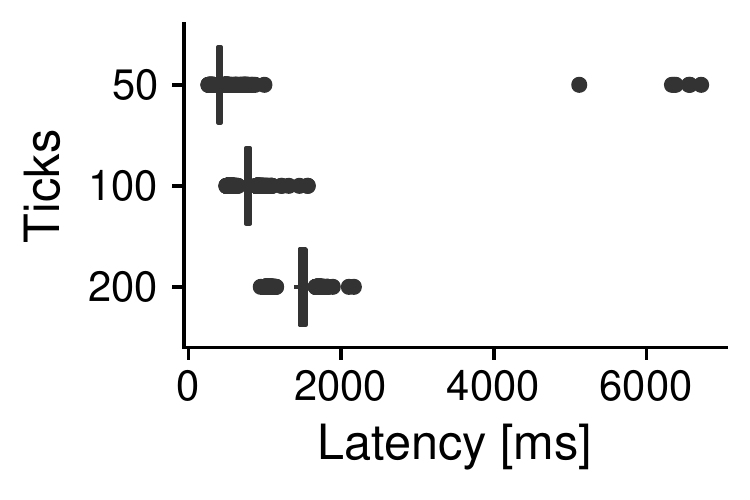}
    \end{subfigure}
    \begin{subfigure}[b]{0.49\linewidth}
        \includegraphics[width=\linewidth]{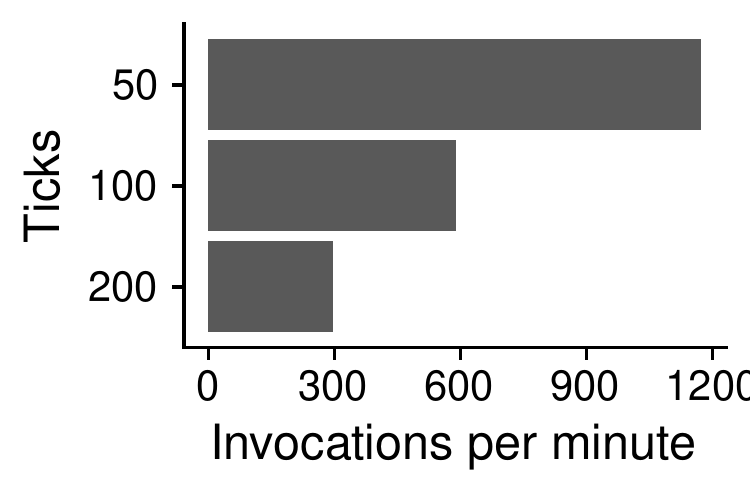}
    \end{subfigure}
    \caption{Latency~(left) and number of invocations per minute~(right) for varying simulations lengths in ticks.}
    \label{fig:res:offloading-latency-invocations}
    \vcutM{}
\end{figure}

The right-hand plot in Figure~\ref{fig:res:offloading-efficiency} shows the function efficiency for a varying number of simulation steps, using a fixed 20-tick lead.
The plot shows that efficiency remains high~(100\% median) for both 50 and 100 simulation steps, but drops below 100\% for 200 steps.
We find the reason for this drop in efficiency is the increased latency of the function when simulating 200 steps, which we explore in Figure~\ref{fig:res:offloading-latency-invocations}.

Figure~\ref{fig:res:offloading-latency-invocations} shows the end-to-end latency of invoking a simulation function, and the number of invocations per minute, for varying simulation lengths.
The left-hand plot shows that increasing the simulation length increases the function latency.
The 200-tick simulation shows a mean latency of 1459\,ms, which exceeds the lead time of 1000\,ms (20 ticks of 50\,ms each).
This explains the reduced efficiency in the right-hand plot in Figure~\ref{fig:res:offloading-efficiency}.

Figures~\ref{fig:res:offloading-efficiency} and~\ref{fig:res:offloading-latency-invocations} show numerous outliers in all plots.
Further analysis shows that outliers have a strong temporal correlation, providing evidence that these outliers are caused by cold starts.
This is surprising because we perform a warm-up iteration of the experiment that uses more invocations per minute than the other configurations,
suggesting that AWS starts deallocating function resources within minutes.

Multiplying the mean latency with the number of invocations per minute shows the additional cost of
running \sys to be between \textdollar0.216 and \textdollar0.244 per hour.
This is comparable to the cost of running one c5n.xlarge instance~(\textdollar0.216 per hour),
but without the instance management overhead.

\subsection{Serverless Terrain Generation Provides Good \qos}
\label{sec:experiments:terrain-gen-qos}
\label{sec:experiments:terrain-gen-aws}

\begin{figure}[t]
    \begin{subfigure}{0.49\linewidth}
        \includegraphics[width=\linewidth]{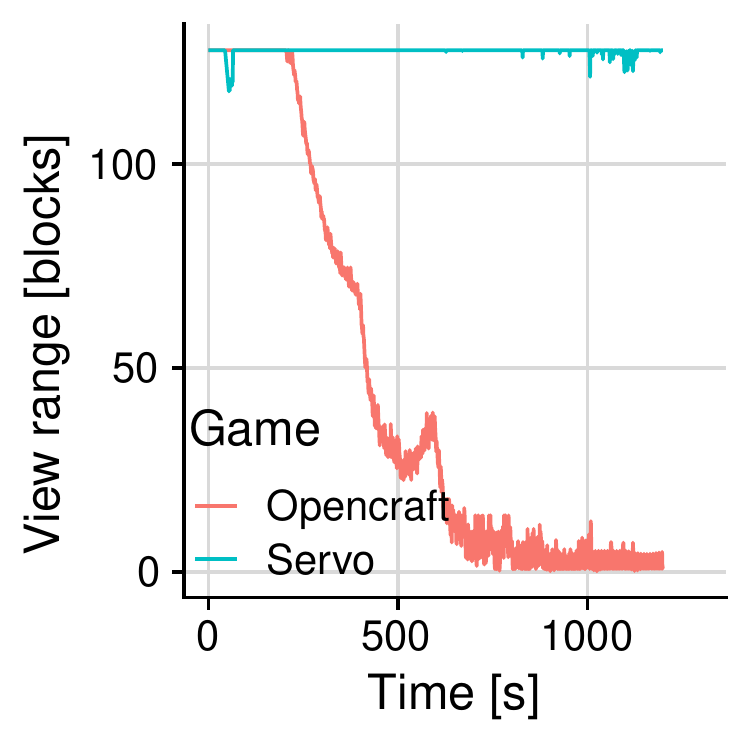}
        \caption{Distance until closest unloaded chunk. Higher is better.}
        \label{fig:exp:terrain-gen:tiancu:distance}
    \end{subfigure}
    \hfill
    \begin{subfigure}{0.49\linewidth}
        \includegraphics[width=\linewidth]{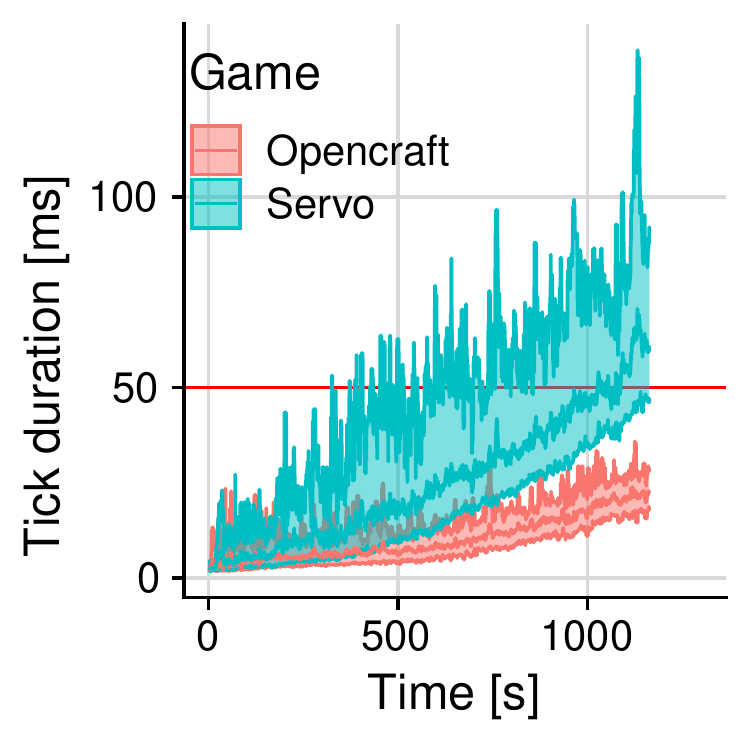}
        \caption{Tick duration over time. Lower is better.}
        \label{fig:exp:terrain-gen:tiancu:tickduration}
    \end{subfigure}
    \caption{Serverless Terrain Generation Performance.}
    \label{fig:exp:terrain-gen-qos}
    \vcutM{}
\end{figure}

Procedurally generating terrain for \mves using serverless functions achieves good \qos.
Figure~\ref{fig:exp:terrain-gen-qos} shows this result.
The left-hand plot shows how \sys is able to load the required terrain in time,
whereas Opencraft does not.
The horizontal axis shows time, and the vertical axis shows the minimum distance between a player and the closest missing part of terrain.
To provide good \qos, this value should remain at 128, the default configured view distance of the players.
The workload consists of five players, each moving in a different direction.
The players initially move at a speed of one unit per second, and increase their speed by an additional unit per second every 200 seconds.

The plots show that Opencraft (red curve) manages to generate sufficient terrain when the players move at 1 unit per second,
but fails to do so when their speed increases.
After 1000 seconds, when the players move 6 units per second, the closest unloaded terrain is less than 16 units away.
I.e., the terrain generation cannot keep up with the workload.
In contrast, \emph{\sys maintains good \qos throughout the experiment}.
After a short performance degradation after roughly 30 seconds (dip in the blue curve in the top left),
the system maintains stable performance.

The right-hand plot shows the tick duration for both games.
The vertical axis shows the tick duration, which must remain below 50\,ms to meet \qos constraints.
The outsides of the colored bands indicate the 5\textsuperscript{th} and 95\textsuperscript{th} percentiles using a 2.5-second window,
and the curve inside the band shows a 2.5-second rolling arithmetic mean.
Although \sys maintains good \qos in terms of terrain generation,
the 95\textsuperscript{th} percentile tick duration starts exceeding the 50\,ms threshold while players move at a speed of 2~blocks per second.
Further analysis shows that this is caused by the higher number of terrain parts generated for \sys compared to Opencraft.
Although content generation happens outside the main game loop for both games,
the overhead of loading the content in the game causes overhead which increases tick duration.

\begin{techrep}
    Functions with few resources result in high latency variability.
    This result complements existing work, which observes increased bandwidth variability in functions with fewer resources~\cite{DBLP:conf/usenix/WangLZRS18}.
    Figure~\ref{fig:res:terrain-gen-aws} shows this result.

    On AWS~Lambda, the amount of compute resources (typically expressed in the number of vCPUs) available to the serverless
    function determined by the amount of memory the user allocates to the function.
    Functions with larger amounts of memory receive more vCPUs.
    Because terrain generation is a compute-intensive task, we expect functions with more memory (and therefore more vCPUs)
    to perform better (i.e., achieve lower latency).

    Figure~\ref{fig:res:terrain-gen-aws:latency} shows the latency of generating a single chunk (an area of $16 \times 16
        \times 256$ blocks) for varying memory configurations on AWS Lambda.
    The results confirm our hypothesis, and show that functions with more memory achieve significantly lower latency.
    The vertical axis shows six different memory configurations for the terrain generation function,
    and the horizontal axis shows the latency, as measured from the MVE server.
    The box whiskers indicate values within 1.5 times the interquartile range, the red dots indicate the arithmetic mean,
    and the values in parentheses are the maximum values. Other outliers are hidden for improved readability.
    A function with 10,240\,MB memory generates a chunk in less than 1~second on average, whereas a function with 320\,MB memory takes
    more than 3~seconds on average.

    Interestingly, performance variability (as indicated by the box width and maximum value) increases for a decreasing amount of resources.
    This incentivizes developers of latency-sensitive applications to select, and pay for,
    more resources than required to avoid worst-case performance outliers.

    Figure~\ref{fig:res:terrain-gen-aws:cost-efficiency} shows that terrain generation performance scales sublinearly with
    the amount of available resources.
    The vertical axis shows six different memory configurations for the terrain generation function,
    and the horizontal axis shows the normalized performance-to-cost ratio,
    which we compute by multiplying the average latency by the amount of memory allocated to the serverless function and applying an inverse normalization function.
    Higher values are better.
    The figure shows that, except for the 320\,MB configuration, functions with fewer resources obtain higher relative
    performance, incentivizing users to reduce the amount of resources allocated to their functions.

    \begin{figure}[t]
        \centering
        \begin{subfigure}[b]{.49\linewidth}
            \includegraphics[width=\linewidth]{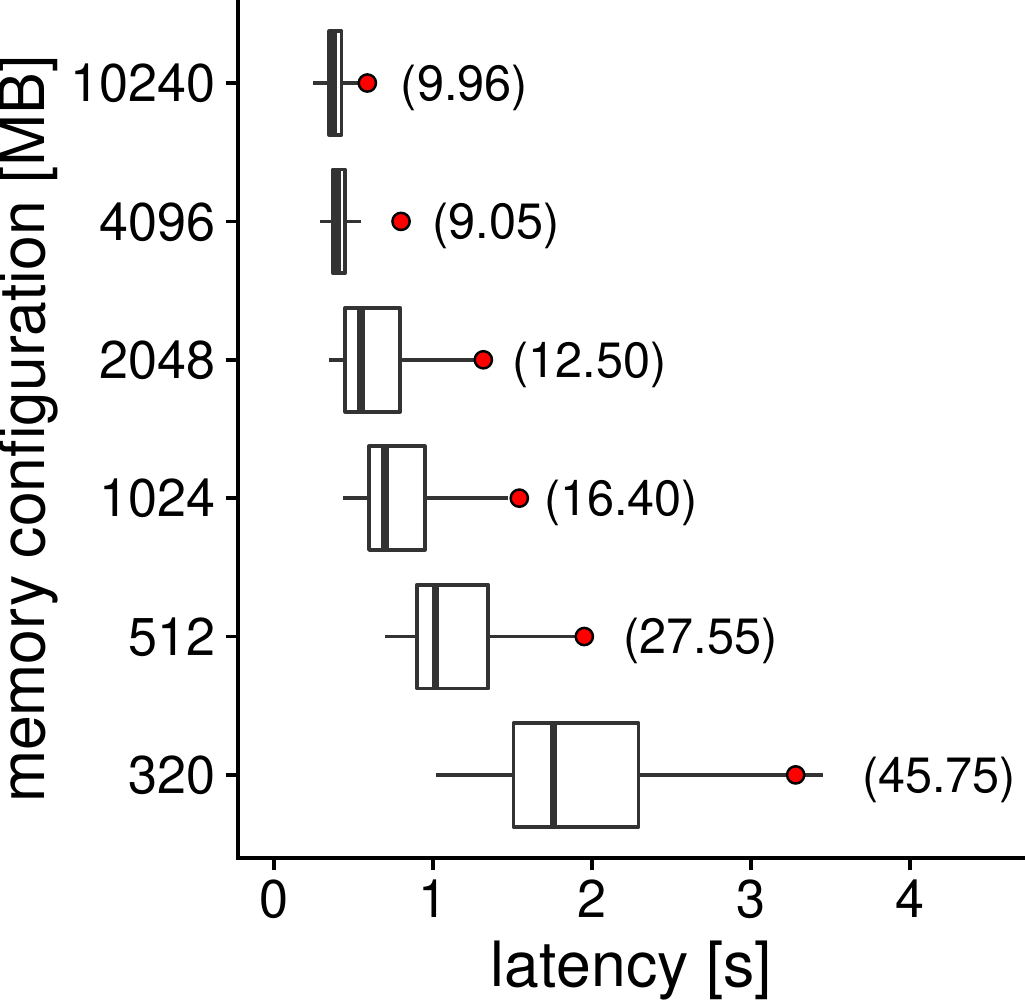}
            \caption{Terrain generation latency.}
            \label{fig:res:terrain-gen-aws:latency}
        \end{subfigure}
        \begin{subfigure}[b]{.49\linewidth}
            \includegraphics[width=\linewidth]{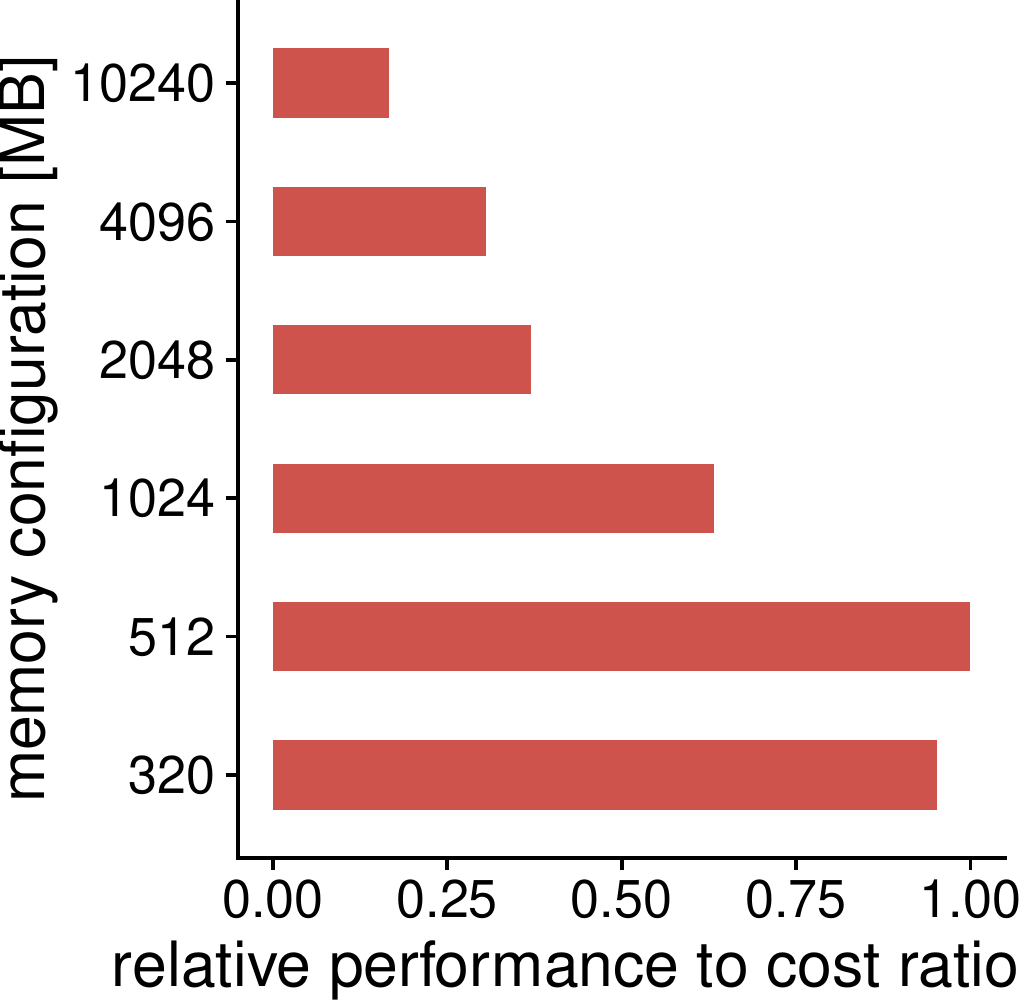}
            \caption{Generation cost-efficiency.}
            \label{fig:res:terrain-gen-aws:cost-efficiency}
        \end{subfigure}
        \caption{Serverless terrain generation on AWS Lambda.}
        \label{fig:res:terrain-gen-aws}
    \end{figure}
\end{techrep}

\subsection{Serverless Terrain Generation Has Good Performance}
\label{sec:experiments:serverless-improves-scalability}
\label{sec:experiments:terrain-gen-perf}

\begin{figure}[!t]
    \centering
    \begin{subfigure}[b]{\linewidth}
        \includegraphics[width=\linewidth]{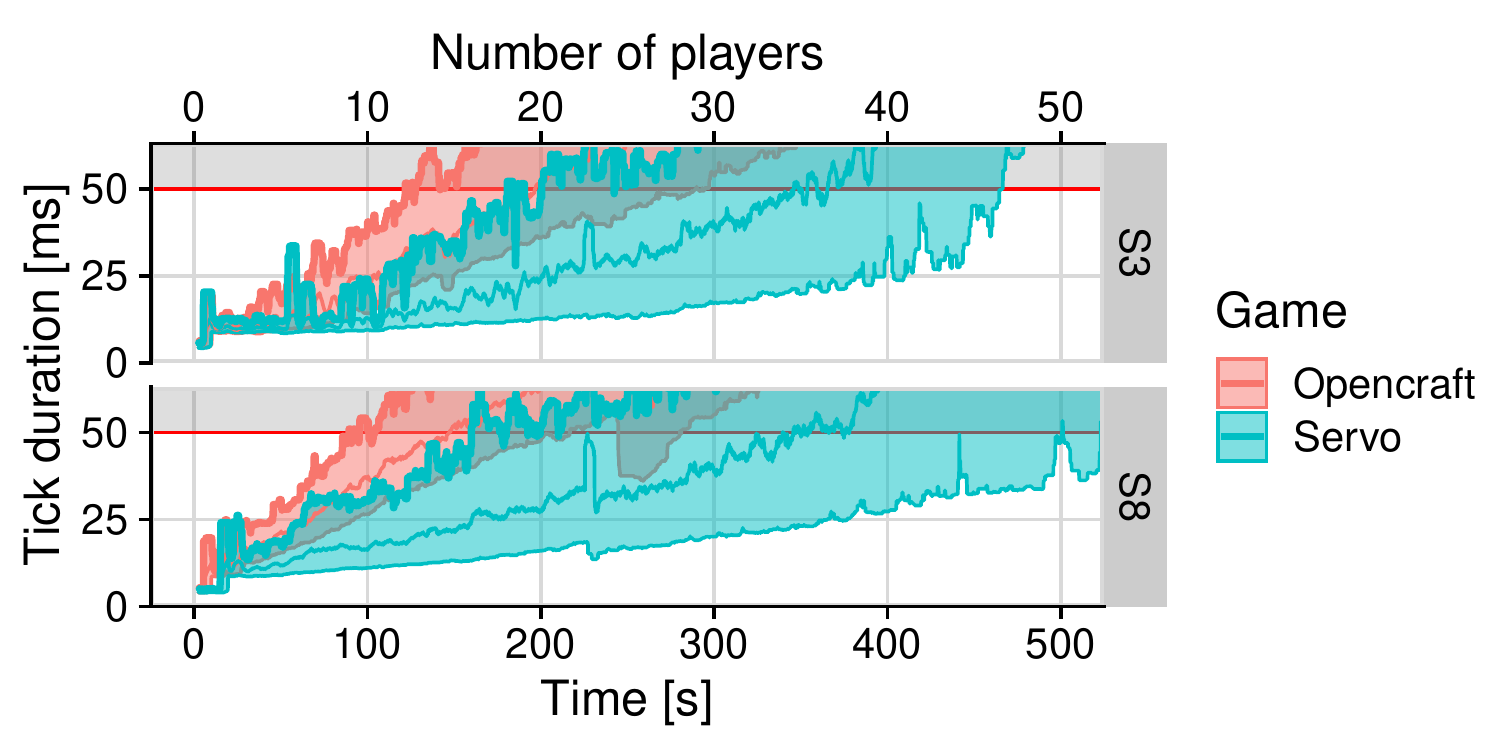}
        \caption{Tick duration over an increasing number of players for workloads \sw3 and \sw8. Lower is better. Values must remain below 50\,ms to maintain good \qos.}
        \label{fig:res:players-azure:s}
    \end{subfigure}
    \begin{subfigure}[b]{\linewidth}
        \includegraphics[width=\linewidth]{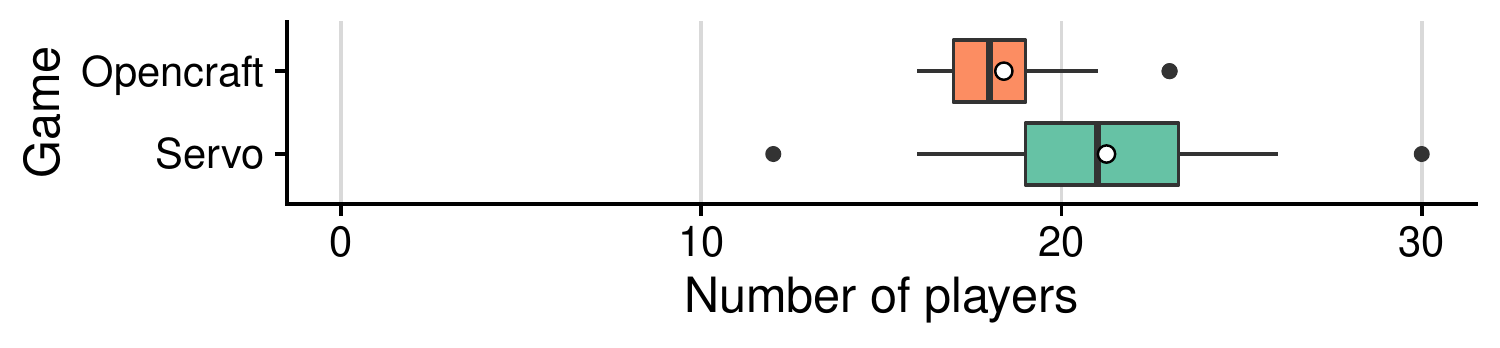}
        \caption{Maximum supported players for the \rw{} workload. Higher is better. The white dot indicates the arithmetic mean.}
        \label{fig:res:players-azure:r}
    \end{subfigure}
    \caption{Performance improvement of serverless terrain generation (\sys) compared to local generation (Opencraft) for varying workloads.}
    \label{fig:res:players-azure}
    \vcutM{}
\end{figure}

Under constant workload, serverless terrain generation provides lower tick times than Opencraft,
increasing the median number of supported players by 17\%, and increasing performance overall.
This section presents experimental evidence for this result, for three separate workloads.

Figure~\ref{fig:res:players-azure:s} shows the tick durations over time for the \sw3 and \sw8 workloads,
in which players move away from the starting location in a straight line with a speed of 3 or 8 in-game units per second, respectively.
Each player goes in a different direction, exploring new areas and generating new terrain.

The horizontal axis on the bottom shows time, and the horizontal axis on the top shows the number of players.
As time progresses, more players connect to the game.
The vertical axis shows the tick duration, which must remain below 50\,ms to meet \qos constraints.
The outsides of the colored bands indicate the 5\textsuperscript{th} and 95\textsuperscript{th} percentiles using a 2.5-second window,
and the curve inside the band shows a 2.5-second rolling arithmetic mean.

The figure shows the tick duration increasing for both Opencraft and \sys.
This is caused by the increasing workload: every ten seconds, a new player joins the game and starts moving in a new direction.
This causes additional terrain to be sent to clients and more terrain to be generated, increasing the load on the server.
Under the \sw3 workload, Opencraft and \sys support up to 12 and 18 players respectively (the 95\textsuperscript{th} percentile curve exceeds the 50\,ms threshold).
The \sw8 workload shows a similar trend, but supports fewer players (9 and 15 respectively) because it is a heavier workload.

Because the plot shows large performance variability (height of colored bands),
we perform an experiment with a random behavior workload which we repeat 20 times.
The result is shown in Figure~\ref{fig:res:players-azure}.
The plot shows that \sys supports slightly more players than Opencraft (distribution is more to the right),
but has increased performance variability (width of the box is larger).
The plot shows that \sys maintains good performance when using \faas to offload terrain generation.

\begin{techrep}
    \subsection{Simple Caching Significantly Reduces Latency Variability}
    \label{sec:experiments:storage-cache-policy}

    \begin{figure}[t]
        \includegraphics[width=\linewidth]{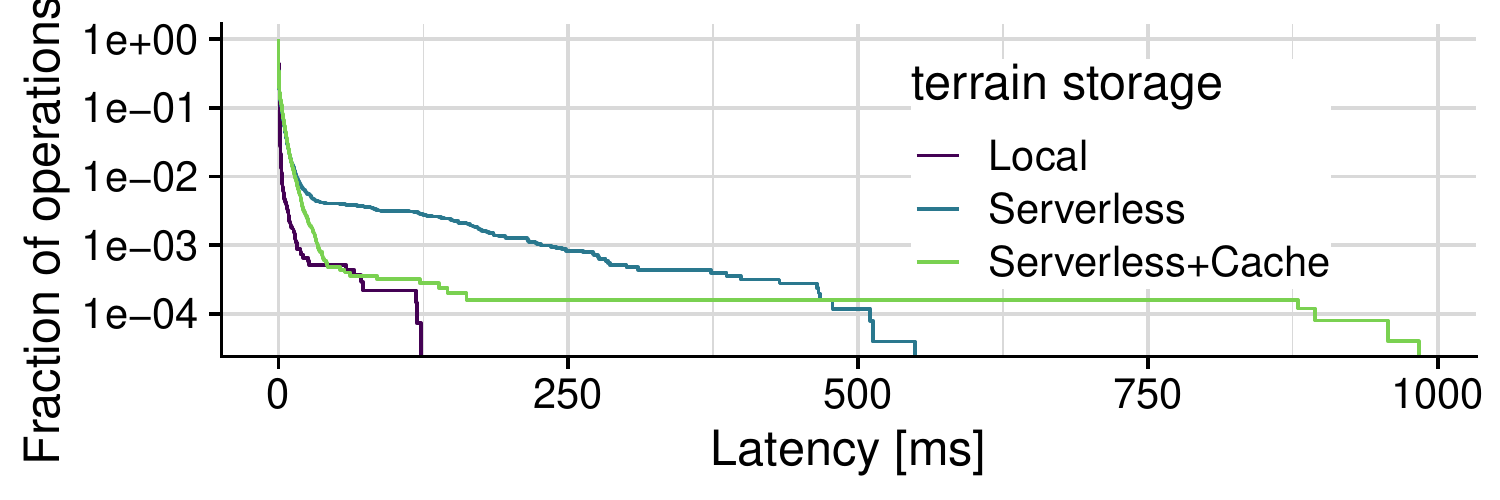}
        \caption{Terrain retrieval latency for local and cloud storage.}
        \label{fig:res:storage-azure-latency-cache-policy}
    \end{figure}

    Caching policies can make cloud storage suitable for use in \mve environments.
    Figure~\ref{fig:res:storage-azure-latency-cache-policy} shows this result.
    The figure shows an inverse cumulative distribution function for terrain retrieval latency.
    The horizontal axis shows the end-to-end latency as observed from the game server,
    and the vertical axis shows the fraction of terrain loading operations, on a logarithmic scale.
    The curves contain between 13682 and 24961 samples each.

    Local storage (dark purple) performs best.
    It completes 99.9\% of requests within 16\,ms,
    and no requests exceed 123\,ms.
    The outliers are visible as two ``shoulders'' in the purple curve.
    Further analysis shows that the outliers occurred during boot,
    all within 4 seconds of starting the game.

    Serverless storage (blue curve) performs significantly worse.
    Its performance variability is significantly higher than local storage,
    which is visible in the plot by the horizontal range of the curve.
    Its 99\textsuperscript{th} percentile latency (16\,ms) is five times higher than local storage,
    and the 99.9\textsuperscript{th} percentile latency (226\,ms), is 14 times worse than local storage.
    Outliers reach 500\,ms latency, which is consistent with the behavior we observed in Figure~\ref{fig:res:storage-azure-latency}.
    Because not receiving terrain data in time can cause visual glitches to players,
    this approach is not suitable for \mves.

    The \emph{simple distance} caching policy (light green curve) generally performs much better than serverless storage without caching, but shows very large outliers close to 984\,ms.
    Although the 99\textsuperscript{th} percentile performance is similar to serverless storage without caching,
    its 99.9\textsuperscript{th} percentile latency is 34\,ms, almost seven times lower than serverless storage without caching, and lower than a single simulation step (50\,ms).
    Surprisingly, the outliers for the approach \emph{with} caching far exceed the outliers for the serverless storage without caching.
    Further analysis shows that there are only seven samples above 500\,ms,
    all of which occur when starting the game.
    Because we run the configuration with caching roughly 15 minutes before the configuration without caching,
    and because of the observations mentioned above, we conjecture that behavior is a cold start phenomenon.

    \subsection{Serverless Offloading for Simulated Constructs has Good Performance for Small- to Medium-Sized Simulated Constructs}
    \label{sec:experiments:offloading-sc-size-performance}
\end{techrep}

\section{Related Work}
\label{sec:related-work}

In this section, we present a developing overview of related work.
The benefits of using serverless services have been studied for several novel application domains such as data processing~\cite{DBLP:conf/cloud/JonasPVSR17}, graph processing~\cite{DBLP:conf/ispdc/ToaderUMI19}, and video encoding~\cite{DBLP:conf/nsdi/FouladiWSBZBSPW17}.
Furthermore, the application of fine-grained computational units and serverless computing for gaming services has been an active research topic for the past decade~\cite{bernstein2014orleans, Bors2023}.
However, \emph{\sys is the first online multiplayer game to improve its virtual-world scalability through serverless computing.}
Supporting real-time online-multiplayer games on serverless platforms is challenging because
real-time games require consistently high performance and low latency,
and because the FaaS programming model (request/reply) does not match well that of real-time online-multiplayer games~(message streams).

There exist a variety of architectures for scalable online multiplayer games.
An extensive overview is available in~\cite{DBLP:journals/corr/abs-2102-09847}.
Systems such as VON~\cite{DBLP:journals/network/HuCC06}, Colyseus~\cite{DBLP:conf/nsdi/BharambePS06}, and Rokkatan~\cite{DBLP:conf/ACMace/MullerMPSG05} distribute the simulation of a single static world over multiple machines.
In contrast, \sys uses additional resources dynamically and is designed specifically to address the computational challenges of \mves.

\begin{techrep}
	VON~\cite{DBLP:journals/network/HuCC06} is a large-scale online gaming system designed to support large numbers of users.
	VON uses a peer-to-peer~(P2P) approach which avoids the use of a central server and lets clients connect to each other directly to avoid a central network bottleneck.

	Colyseus~\cite{DBLP:conf/nsdi/BharambePS06} in a P2P architecture for large-scale online-multiplayer games.
	It supports a large number of players by partitioning game objects (e.g., avatars, items) over the participating nodes using a distributed hash table~(DHT),
	and pre-fetching these objects to hide latency.

	Rokkatan~\cite{DBLP:conf/ACMace/MullerMPSG05} is a large-scale online game that uses multi-server replication to improve scalability.
	This approach is similar to a regular client/server architecture, but instead partitions objects across the available servers, and places read-only copies of these objects on all other servers.

	However, these architectures do not address important challenges present in \mves:
	on-demand terrain generation and the simulation of a modifiable world and embedded \emph{simulated constructs}.
	\sys addresses these challenges using FaaS for computational offloading and speculative execution.

\end{techrep}

Speculative execution techniques have been applied in a variety of domains,
such as big data processing~\cite{DBLP:conf/icdcs/XuALS18}, state machine
replication~\cite{DBLP:conf/nsdi/WesterCNCFL09}, and in a variety of interactive application types:
VNC/SRD~\cite{DBLP:conf/usenix/LangeDR08} uses speculative execution to predict frames for a remote display system.
Crom~\cite{DBLP:conf/nsdi/MickensEHL10} performs speculative execution to hide latency in Web browsing.
Outatime~\cite{DBLP:conf/mobisys/LeeCCKDGWF15} uses speculative execution to render frames based on a player's predicted
inputs.
Although speculative execution is commonly applied close to the user,
\sys uses speculative execution for computational offloading.

Mirror~\cite{DBLP:journals/concurrency/JiangVPI18} is an architecture for computation-offloading in mobile games, in
which mobile games can offload computation to a remote machine.
Whereas Mirror requires the remote machine to be always online, and delays player inputs to hide latency,
\sys uses serverless functions that only run when the offloaded computation is required and
hides latency by performing local active replication until the offloaded simulation is complete.

\begin{techrep}
	\emph{Symmetric active-active replication}~\cite{DBLP:conf/cluster/UhlemannES06} is a process replication technique from the field of high-performance computing in which multiple instances of a
	service receive the same inputs and perform the same operations to hide failures from users and improve availability.
	Similarly, the SC offloading technique presented in this work also uses two active simulators with the same input.
	However, in contrast to symmetric active-active replication, our technique focuses on hiding latency (not failures).
	Moreover, it is not strictly symmetrical because one of the two services (the serverless simulator) performs speculative
	execution, and does not process player input.
\end{techrep}

\section{Conclusion and Future Work}
\label{sec:conclusion}
\label{sec:future-work}

Gaming is the world's largest entertainment industry, and
\mves are the best-selling game of all time.
Minecraft, the canonical example of an \mve, has more than 130~million active monthly players.
However, due to the inherent complexity of \emph{modifiable} virtual worlds,
this scale can only be reached by replicating large numbers of isolated instances,
preventing players from playing together.
In this work, we explore the use of serverless computing to address the complexity and limited scalability of \mve
instances.
Specifically, we design \sys, a prototype \mve which leverages serverless to offload both compute and storage from \mve
server instances to the cloud.
Our results show that \sys's offloading increases the number of supported players by 140,
and that it can successfully hide (cold-start) latency from players.

In future work, we aim to investigate how to scale large simulated constructs while meeting their \qos requirements, and (serverless) \mve designs without a central server.

\bibliographystyle{IEEEtran}
\bibliography{references_beautified}

\begin{thebibliography}{10}
\providecommand{\url}[1]{#1}
\csname url@samestyle\endcsname
\providecommand{\newblock}{\relax}
\providecommand{\bibinfo}[2]{#2}
\providecommand{\BIBentrySTDinterwordspacing}{\spaceskip=0pt\relax}
\providecommand{\BIBentryALTinterwordstretchfactor}{4}
\providecommand{\BIBentryALTinterwordspacing}{\spaceskip=\fontdimen2\font plus
\BIBentryALTinterwordstretchfactor\fontdimen3\font minus
  \fontdimen4\font\relax}
\providecommand{\BIBforeignlanguage}[2]{{%
\expandafter\ifx\csname l@#1\endcsname\relax
\typeout{** WARNING: IEEEtran.bst: No hyphenation pattern has been}%
\typeout{** loaded for the language `#1'. Using the pattern for}%
\typeout{** the default language instead.}%
\else
\language=\csname l@#1\endcsname
\fi
#2}}
\providecommand{\BIBdecl}{\relax}
\BIBdecl

\bibitem{DBLP:conf/fdg/Bar-ElR20}
D.~Bar{-}El and K.~E. Ringland, ``Crafting game-based learning: An analysis of
  lessons for minecraft education edition,'' in \emph{{FDG}}, 2020.

\bibitem{Melchiorre2022}
{Michele Melchiorre, Senior VP BMW}, ``The {BMW} i{F}actory - {H}ighly
  {E}fficient and {C}ompetitive,'' in \emph{Keynote ISC}, 2022.

\bibitem{MinecraftOfficiallyCrosses2021}
``Minecraft officially crosses over 141 million monthly active users,''
  https://www.windowscentral.com/minecraft-live-2021-numbers-update, October
  2021.

\bibitem{DBLP:conf/wosp/SarDI19}
{van der Sar} \emph{et~al.}, ``Yardstick: {{A}} benchmark for minecraft-like
  services,'' in \emph{ICPE}, 2019.

\bibitem{MinecraftRealms2022Jan}
\BIBentryALTinterwordspacing
Mojang, ``{Minecraft Realms},'' Jan. 2022, ``You and up to 10 friends can play
  together at one time.''. [Online]. Available:
  \url{https://www.minecraft.net/en-us/realms}
\BIBentrySTDinterwordspacing

\bibitem{book:system:Dear17PLATO}
B.~Dear, Ed., \emph{The Friendly Orange Glow: The Untold Story of the PLATO
  System and the Dawn of Cyberculture}.\hskip 1em plus 0.5em minus 0.4em\relax
  Pantheon Books, 2017.

\bibitem{newzoo2021}
{Newzoo}, ``Newzoo global games market report 2021 free version newzoo,'' July
  2021.

\bibitem{DBLP:conf/hotcloud/DonkervlietTI20}
J.~Donkervliet \emph{et~al.}, ``Towards supporting millions of users in
  modifiable virtual environments by redesigning minecraft-like games as
  serverless systems,'' in \emph{HotCloud}, 2020.

\bibitem{DBLP:journals/internet/EykTTVUI18}
Eyk \emph{et~al.}, ``Serverless is more: {{From PaaS}} to present cloud
  computing,'' \emph{IC}, vol.~22, 2018.

\bibitem{DBLP:journals/cacm/Schleier-SmithS21}
J.~Schleier{-}Smith \emph{et~al.}, ``What serverless computing is and should
  become: the next phase of cloud computing,'' \emph{Commun. {ACM}}, vol.~64,
  no.~5, 2021.

\bibitem{DBLP:journals/cacm/CastroIMS19}
P.~C. Castro \emph{et~al.}, ``The rise of serverless computing,'' \emph{Commun.
  {ACM}}, vol.~62, no.~12, 2019.

\bibitem{DBLP:conf/icdcs/DonkervlietCI21}
Donkervliet \emph{et~al.}, ``Dyconits: {{Scaling}} minecraft-like services
  through dynamically managed inconsistency,'' in \emph{ICDCS}, 2021.

\bibitem{DBLP:journals/corr/abs-2102-09847}
Gonz{\'a}lez \emph{et~al.}, ``Key technologies for networked virtual
  environments: {{A}} new taxonomy,'' \emph{CoRR}, vol. abs/2102.09847, 2021.

\bibitem{DBLP:journals/csur/LiuBC12}
Liu \emph{et~al.}, ``Survey of state melding in virtual worlds,'' \emph{ACM
  Comput. Surv.}, vol.~44, 2012.

\bibitem{DBLP:journals/csur/LiuT14}
Liu and Theodoropoulos, ``Interest management for distributed virtual
  environments: {{A}} survey,'' \emph{ACM Comput. Surv.}, vol.~46, 2014.

\bibitem{DBLP:journals/tpds/GilmoreE12}
Gilmore and Engelbrecht, ``A {{Survey}} of {{State Persistency}} in
  {{Peer-to-Peer Massively Multiplayer Online Games}},'' \emph{TPDS}, vol.~23,
  2012.

\bibitem{DBLP:conf/hci/WorsleyTMZJ21}
Worsley \emph{et~al.}, ``Multicraft: {{A}} multimodal interface for supporting
  and studying learning in minecraft,'' in \emph{HCI-Games}, vol. 12790, 2021.

\bibitem{DBLP:conf/chi/SlovakSTF18}
Slov{\'a}k \emph{et~al.}, ``Mediating conflicts in minecraft: {{Empowering}}
  learning in online multiplayer games,'' in \emph{CHI}, 2018.

\bibitem{DBLP:conf/acmidc/ZhuH17}
Zhu and Heun, ``Teaching and learning of chinese history in minecraft: {{A}}
  pilot case-study in hong kong secondary schools,'' in \emph{IDC}, 2017.

\bibitem{MinecraftOfficialSite}
``Minecraft {{Official Site}} | {{Minecraft Education Edition}},''
  https://education.minecraft.net/en-us/homepage.

\bibitem{Natividad2018Jan}
\BIBentryALTinterwordspacing
A.~Natividad, ``{How Greenpeace Used Minecraft to Stop Illegal Logging in
  Europe{'}s Last Lowland Primeval Forest},'' \emph{Adweek}, Jan. 2018.
  [Online]. Available: \url{https://bit.ly/MinecraftGreenpeace}
\BIBentrySTDinterwordspacing

\bibitem{UncensoredLibraryReporters}
\BIBentryALTinterwordspacing
``The {{Uncensored Library}} \textendash{} {{Reporters}} without borders.''
  [Online]. Available: \url{https://www.uncensoredlibrary.com/}
\BIBentrySTDinterwordspacing

\bibitem{mvesOnSteam}
\BIBentryALTinterwordspacing
``{Steam Search},'' Jan 2021. [Online]. Available:
  \url{http://bit.ly/steam-mves}
\BIBentrySTDinterwordspacing

\bibitem{mcmods}
\BIBentryALTinterwordspacing
``{Mods - Minecraft - CurseForge},'' Jan 2021. [Online]. Available:
  \url{http://bit.ly/ModsForMinecraft}
\BIBentrySTDinterwordspacing

\bibitem{Meta2022}
J.~Mann, ``{Meta's spent {\ifmmode\$\else\textdollar\fi}36 billion on the
  metaverse. The iPhone and Xbox cost way less.}'' October 2022.

\bibitem{book:metaverse:ShowCrash92}
N.~Stephenson, Ed., \emph{Snow Crash}.\hskip 1em plus 0.5em minus 0.4em\relax
  Bantam Spectra Books, 1992.

\bibitem{DBLP:conf/wosp/EykSEAI20}
Eyk \emph{et~al.}, ``Beyond microbenchmarks: {{The SPEC-RG}} vision for a
  comprehensive serverless benchmark,'' in \emph{ICPE Companion}, 2020.

\bibitem{DBLP:conf/cloud/YuLDXZLYQ020}
Yu \emph{et~al.}, ``Characterizing serverless platforms with serverlessbench,''
  in \emph{SoCC}, 2020.

\bibitem{DBLP:conf/mmsys/RaaenP15}
K.~Raaen and A.~Petlund, ``How much delay is there really in current games?''
  in \emph{MMSys}, 2015.

\bibitem{DBLP:conf/chi/IvkovicSGS15}
Z.~Ivkovic \emph{et~al.}, ``Quantifying and mitigating the negative effects of
  local latencies on aiming in 3d shooter games,'' in \emph{CHI}, 2015.

\bibitem{DBLP:journals/tpds/NaeIP11}
Nae \emph{et~al.}, ``Dynamic {{Resource Provisioning}} in {{Massively
  Multiplayer Online Games}},'' \emph{TPDS}, vol.~22, 2011.

\bibitem{DBLP:journals/ieeemm/MacedoniaZ97}
Macedonia and Zyda, ``A taxonomy for networked virtual environments,''
  \emph{IEEE Multim.}, vol.~4, 1997.

\bibitem{kounev_et_al:DagRep.11.4.34:ServerlessNotion}
S.~Kounev \emph{et~al.}, ``{Toward a Definition for Serverless Computing},'' in
  \emph{{Serverless Computing (Dagstuhl Seminar 21201)}}, 2021, vol.~11.

\bibitem{DBLP:conf/sc/NaeIPPEF08}
V.~Nae \emph{et~al.}, ``Efficient management of data center resources for
  massively multiplayer online games,'' in \emph{SC}, 2008.

\bibitem{DBLP:journals/cacm/ClaypoolC06}
M.~Claypool and K.~T. Claypool, ``Latency and player actions in online games,''
  \emph{Commun. {ACM}}, vol.~49, no.~11, 2006.

\bibitem{DBLP:journals/computer/BalELNRSSW16}
Bal \emph{et~al.}, ``A medium-scale distributed system for computer science
  research: {{Infrastructure}} for the long term,'' \emph{Computer}, vol.~49,
  2016.

\bibitem{DBLP:conf/usenix/WangLZRS18}
Wang \emph{et~al.}, ``Peeking {{Behind}} the {{Curtains}} of {{Serverless
  Platforms}},'' in \emph{ATC}, 2018.

\bibitem{DBLP:conf/cloud/JonasPVSR17}
Jonas \emph{et~al.}, ``Occupy the cloud: Distributed computing for the 99\%,''
  in \emph{SoCC}, 2017.

\bibitem{DBLP:conf/ispdc/ToaderUMI19}
Toader \emph{et~al.}, ``Graphless: {{Toward Serverless Graph Processing}},'' in
  \emph{ISPDC}, 2019.

\bibitem{DBLP:conf/nsdi/FouladiWSBZBSPW17}
Fouladi \emph{et~al.}, ``Encoding, fast and slow: {{Low-latency}} video
  processing using thousands of tiny threads,'' in \emph{NSDI}, 2017.

\bibitem{bernstein2014orleans}
Bernstein \emph{et~al.}, ``Orleans: {{Distributed}} virtual actors for
  programmability and scalability,'' Tech. Rep., 2014.

\bibitem{Bors2023}
B.~Bors, \emph{Game Backend Development}.\hskip 1em plus 0.5em minus
  0.4em\relax Apress, 2023.

\bibitem{DBLP:journals/network/HuCC06}
Hu \emph{et~al.}, ``{{VON}}: A scalable peer-to-peer network for virtual
  environments,'' \emph{IEEE Netw.}, vol.~20, 2006.

\bibitem{DBLP:conf/nsdi/BharambePS06}
Bharambe \emph{et~al.}, ``Colyseus: {{A}} distributed architecture for online
  multiplayer games,'' in \emph{NSDI}, 2006.

\bibitem{DBLP:conf/ACMace/MullerMPSG05}
M{\"u}ller \emph{et~al.}, ``Rokkatan: Scaling an {{RTS}} game design to the
  massively multiplayer realm,'' in \emph{ACE}, 2005.

\bibitem{DBLP:conf/icdcs/XuALS18}
Xu \emph{et~al.}, ``Chronos: {{A}} unifying optimization framework for
  speculative execution of deadline-critical {{MapReduce}} jobs,'' in
  \emph{ICDCS}, 2018.

\bibitem{DBLP:conf/nsdi/WesterCNCFL09}
Wester \emph{et~al.}, ``Tolerating latency in replicated state machines through
  client speculation,'' in \emph{NSDI}, 2009.

\bibitem{DBLP:conf/usenix/LangeDR08}
Lange \emph{et~al.}, ``Experiences with client-based speculative remote
  display,'' in \emph{ATC}, 2008.

\bibitem{DBLP:conf/nsdi/MickensEHL10}
Mickens \emph{et~al.}, ``Crom: {{Faster}} web browsing using speculative
  execution,'' in \emph{NSDI}, 2010.

\bibitem{DBLP:conf/mobisys/LeeCCKDGWF15}
Lee \emph{et~al.}, ``Outatime: {{Using}} speculation to enable low-latency
  continuous interaction for mobile cloud gaming,'' in \emph{MobiSys}, 2015.

\bibitem{DBLP:journals/concurrency/JiangVPI18}
Jiang \emph{et~al.}, ``A mirroring architecture for sophisticated mobile games
  using computation-offloading,'' \emph{Concurr. Comput. Pract. Exp.}, vol.~30,
  2018.

\bibitem{DBLP:conf/cluster/UhlemannES06}
Uhlemann \emph{et~al.}, ``{{JOSHUA}}: {{Symmetric Active}}/{{Active}}
  replication for highly available {{HPC}} job and resource management,'' in
  \emph{CLUSTER}, 2006.

\end{thebibliography}

\end{document}